\documentclass[twocolumn]{aastex631}
\usepackage{verbatim}
\usepackage{hyperref}
\hypersetup{
    colorlinks=true,
    linkcolor=blue,
    filecolor=magenta,      
    urlcolor=cyan,
    citecolor=blue,
}
\usepackage{amsmath}
\usepackage{comment}
\usepackage{graphicx}
\usepackage{xcolor}

\newcommand{\target}{TOI-3261}

\newcommand{\C}{\ensuremath{^{\circ}C\;}}
\newcommand{\EBV}{E(\ensuremath{B-V})}

\newcommand{\arstar}{\ensuremath{a/\rstar}}

\newcommand{\feh}{[Fe/H]}

\newcommand{\gcmc}{\ensuremath{\rm g\,cm^{-3}}}

\newcommand{\kms}{\ensuremath{\rm km\,s^{-1}}}

\newcommand{\loggstar}{\ensuremath{\log{g_{\star}}}}
\newcommand{\logg}{\ensuremath{\log g}}

\newcommand{\lstar}{\ensuremath{L_\star}}
\newcommand{\lsun}{\ensuremath{L_\sun}}
\newcommand{\masyr}{\ensuremath{\rm mas\,yr^{-1}}}

\newcommand{\mearth}{\ensuremath{M_\earth}}
\newcommand{\mjup}{\ensuremath{M_{\rm Jup}}}
\newcommand{\mpl}{\ensuremath{M_{p}}}

\newcommand{\mstar}{\ensuremath{M_\star}}
\newcommand{\msun}{\ensuremath{M_\sun}}

\newcommand{\pmra}{\ensuremath{\rm \mu_{\alpha}}}
\newcommand{\pmdec}{\ensuremath{\rm \mu_{\delta}}}

\newcommand{\Prot}{\ensuremath{P_{rot}}}
\newcommand{\rearth}{\ensuremath{R_\earth}}

\newcommand{\rhostar}{\ensuremath{\rho_\star}}

\newcommand{\rpl}{\ensuremath{R_{p}}}
\newcommand{\rstar}{\ensuremath{R_\star}}
\newcommand{\rsun}{\ensuremath{R_\sun}}

\newcommand{\teffstar}{\ensuremath{T_{\rm eff\star}}}
\newcommand{\teff}{T$_{\rm eff}$}

\newcommand{\tess}{{\it TESS}}
\newcommand{\vsini}{$v$sin($i$)}

\newcommand{\ticMass}{\ensuremath{0.87}}
\newcommand{\ticRadius}{\ensuremath{0.86}}

\newcommand{\starDistance}{\ensuremath{300.5\pm1.1}}
\newcommand{\starMass}{\ensuremath{0.861_{-0.030}^{+0.020}}}
\newcommand{\starRadius}{\ensuremath{0.849_{-0.017}^{+0.024}}}
\newcommand{\starLuminosity}{\ensuremath{0.430_{-.019}^{+0.025}}}
\newcommand{\starTeff}{\ensuremath{5068_{-42}^{+61}}}
\newcommand{\starfeh}{\ensuremath{0.138_{-0.037}^{+0.048}}}
\newcommand{\starTESSuOne}{\ensuremath{0.59_{-0.39}^{+0.53}}}
\newcommand{\starTESSuTwo}{\ensuremath{0.06_{-0.44}^{+0.28}}}
\newcommand{\starLCOuOne}{\ensuremath{0.35_{-0.19}^{+0.30}}}
\newcommand{\starLCOuTwo}{\ensuremath{0.04_{-0.23}^{+0.25}}}

\newcommand{\bSemiAmplitude}{\ensuremath{22.1_{-1.5}^{+1.9}}}

\newcommand{\jitterESPRESSO}{\ensuremath{3.8_{-1.1}^{+1.0}}}
\newcommand{\jitterHARPS}{\ensuremath{5.2_{-3.1}^{+3.8}}}
\newcommand{\gammaESPRESSO}{\ensuremath{-1910.2_{-1.4}^{+1.3}}}
\newcommand{\gammaHARPS}{\ensuremath{-1921.8_{-2.6}^{+1.8}}}

\newcommand{\starRho}{\ensuremath{1.400_{-0.098}^{+0.077}}}
\newcommand{\starLogg}{\ensuremath{4.19\pm0.123}}
\newcommand{\starTESSqOne}{\ensuremath{0.43_{-0.24}^{+0.39}}}
\newcommand{\starTESSqTwo}{\ensuremath{0.45_{-0.22}^{+0.36}}}
\newcommand{\starLCOqOne}{\ensuremath{0.18_{-0.12}^{+0.24}}}
\newcommand{\starLCOqTwo}{\ensuremath{0.31_{-0.23}^{+0.34}}}

\newcommand{\bEpoch}{\ensuremath{2459112.40036_{-0.00101}^{+0.00063}}}
\newcommand{\bPeriod}{\ensuremath{0.8831331_{-0.0000010}^{+0.0000010}}}
\newcommand{\bROR}{\ensuremath{0.0411_{-0.0032}^{+0.0046}}}
\newcommand{\bMass}{\ensuremath{30.3_{-2.4}^{+2.2}}}
\newcommand{\bEcc}{\ensuremath{0}}
\newcommand{\bDensity}{\ensuremath{3.0_{-0.8}^{+1.1}}}
\newcommand{\bomega}{\ensuremath{0}}
\newcommand{\bImpactParameter}{\ensuremath{0.17_{-0.14}^{+0.11}}}
\newcommand{\bSemimajorAxis}{\ensuremath{0.01714_{-0.00020}^{+0.00013}}}
\newcommand{\bAOR}{\ensuremath{4.333_{-0.104}^{+0.078}}}
\newcommand{\bInclination}{\ensuremath{87.8_{-1.5}^{+1.8}}}
\newcommand{\bRadius}{\ensuremath{3.82_{-0.35}^{+0.42}}}
\newcommand{\bTeq}{\ensuremath{1722_{-18}^{+26}}}
\newcommand{\bIrr}{\ensuremath{1466_{-60}^{+92}}}
\newcommand{\bDuration}{\ensuremath{1.618_{-0.043}^{+0.024}}}
\newcommand{\bIngressDuration}{\ensuremath{3.96_{-0.33}^{+0.46}}}

\newcommand{\bApproximatePeriod}{\ensuremath{0.88}}

\newcommand{\bRM}{\ensuremath{1}}

\shorttitle{The Discovery of TOI-3261b}
\shortauthors{Nabbie et al.}

\begin{document}

\title{Surviving in the Hot Neptune Desert: The Discovery of the Ultra-Hot Neptune TOI-3261b}

\author[0000-0003-0571-2245]{Emma Nabbie}
\affiliation{University of Southern Queensland, West St, Darling Heights, Toowoomba, Queensland, 4350, Australia}
\author[0000-0003-0918-7484]{Chelsea~X.~Huang}
\affiliation{University of Southern Queensland, West St, Darling Heights, Toowoomba, Queensland, 4350, Australia}
\author[0000-0002-0040-6815]{Jennifer A. Burt}
\affiliation{Jet Propulsion Laboratory, California Institute of Technology, 4800 Oak Grove Drive, Pasadena, CA 91109, USA}
\author[0000-0002-5080-4117]{David J. Armstrong}
\affiliation{Department of Physics, University of Warwick, Gibbet Hill Road, Coventry CV4 7AL, UK}
\affiliation{Centre for Exoplanets and Habitability, University of Warwick, Gibbet Hill Road, Coventry CV4 7AL, UK}
\author[0000-0003-2008-1488]{Eric E. Mamajek}
\affiliation{Jet Propulsion Laboratory, California Institute of Technology, 4800 Oak Grove Drive, Pasadena, CA 91109, USA}
\author[0000-0002-0601-6199]{Vardan Adibekyan}
\affiliation{Instituto de Astrof{\'i}sica e Ci{\^e}ncias do Espa\c{c}o, Universidade do Porto, CAUP, Rua das Estrelas, 4150-762 Porto, Portugal}
\author[0000-0001-9047-2965]{S\'{e}rgio G. Sousa}
\affiliation{Instituto de Astrof{\'i}sica e Ci{\^e}ncias do Espa\c{c}o, Universidade do Porto, CAUP, Rua das Estrelas, 4150-762 Porto, Portugal}
\author{Eric D. Lopez}
\affiliation{SUPA, Institute for Astronomy, Royal Observatory Edinburgh, University of Edinburgh, Blackford Hill, Edinburgh EH9 3HJ, UK}
\affiliation{NASA Goddard Space Flight Center, 8800 Greenbelt Rd, Greenbelt, MD 20771, USA}
\author[0000-0002-5113-8558]{Daniel Thorngren}
\affiliation{Department of Physics \& Astronomy, Johns Hopkins University, Baltimore, MD, USA}
\author[0000-0002-1416-2188]{Jorge Fern\'{a}ndez Fern\'{a}ndez}
\affiliation{Department of Physics, University of Warwick, Gibbet Hill Road, Coventry CV4 7AL, UK}
\affiliation{Centre for Exoplanets and Habitability, University of Warwick, Gibbet Hill Road, Coventry CV4 7AL, UK}
\author[0000-0001-8308-0808]{Gongjie Li}
\affiliation{Center for Relativistic Astrophysics, School of Physics, Georgia Institute of Technology, Atlanta, GA 30332, USA}
\author[0000-0003-2733-8725]{James S. Jenkins}
\affiliation{Instituto de Estudios Astrof\'{i}sicos, Facultad de Ingenier\'{i}a y Ciencias, Universidad Diego Portales, Av. Ej\'{e}rcito 441, Santiago, Chile}
\affiliation{Centro de Astrof\'{i}sica y Tecnolog\'{i}as Afines (CATA), Casilla 36-D, Santiago, Chile}
\author[0000-0002-1896-2377]{Jose I. Vines}
\affiliation{Instituto de Astronom\'{i}a, Universidad Cat\'{o}lica del Norte, Angamos 0610, 1270709, Antofagasta, Chile}
\author[0000-0001-8056-9202]{Jo\~ao Gomes da Silva}
\affiliation{Instituto de Astrof{\'i}sica e Ci{\^e}ncias do Espa\c{c}o, Universidade do Porto, CAUP, Rua das Estrelas, 4150-762 Porto, Portugal}
\author[0000-0001-9957-9304]{Robert A. Wittenmyer}
\affiliation{University of Southern Queensland, West St, Darling Heights, Toowoomba, Queensland, 4350, Australia}

\author[0000-0001-6023-1335]{Daniel Bayliss}
\affiliation{Department of Physics, University of Warwick, Gibbet Hill Road, Coventry CV4 7AL, UK}
\author[0000-0001-7124-4094]{C\'{e}sar Brice\~{n}o}
\affiliation{Cerro Tololo Inter-American Observatory, Casilla 603, La Serena, Chile}
\author[0000-0001-6588-9574]{Karen A. Collins}
\affiliation{Harvard-Smithsonian Center for Astrophysics, 60 Garden Street, Cambridge, MA 02138, USA}
\author[0000-0002-9332-2011]{Xavier Dumusque}
\affiliation{Observatoire Astronomique de l’Universit\'{e} de Gen\`{e}ve, Chemin Pegasi 51, 1290 Versoix, Switzerland}
\author[0000-0003-1728-0304]{Keith Horne}
\affiliation{SUPA Physics and Astronomy, University of St. Andrews, Fife, KY16 9SS Scotland, UK}
\author[0009-0005-2761-9190]{Marcelo Aron F. Keniger}
\affiliation{Department of Physics, University of Warwick, Gibbet Hill Road, Coventry CV4 7AL, UK}
\affiliation{Centre for Exoplanets and Habitability, University of Warwick, Gibbet Hill Road, Coventry CV4 7AL, UK}
\author{Nicholas Law}
\affiliation{Department of Physics and Astronomy, The University of North Carolina at Chapel Hill, Chapel Hill, NC 27599-3255, USA}
\author[0000-0003-3742-1987]{Jorge	Lillo-Box}
\affiliation{Centro de Astrobiolog\'ia (CAB), CSIC-INTA, Camino Bajo del Castillo s/n, 28692, Villanueva de la Ca\~nada (Madrid), Spain}
\author[0000-0002-9442-137X]{Shang-Fei Liu}
\affiliation{School of Physics and Astronomy, Sun Yat-sen University, Zhuhai, People's Republic of China}
\author[0000-0003-3654-1602]{Andrew W. Mann}
\affiliation{Department of Physics and Astronomy, The University of North Carolina at Chapel Hill, Chapel Hill, NC 27599-3255, USA}
\author[0000-0002-5254-2499]{Louise D. Nielsen}
\affiliation{University Observatory Munich, Ludwig Maximilian University, Scheinerstrasse 1, Munich 81679, Germany}
\author[0000-0002-5899-7750]{Ares Osborn}
\affiliation{Department of Physics and Astronomy, McMaster University, 1280 Main St W, Hamilton, ON, L8S 4L8, Canada}
\author{Howard M. Relles}
\affiliation{Harvard-Smithsonian Center for Astrophysics, 60 Garden Street, Cambridge, MA 02138, USA}
\author[0000-0001-5164-3602]{Jos\'{e} J. Rodrigues}
\affiliation{Instituto de Astrof{\'i}sica e Ci{\^e}ncias do Espa\c{c}o, Universidade do Porto, CAUP, Rua das Estrelas, 4150-762 Porto, Portugal}
\author[0000-0002-8397-557X]{Juan Serrano Bell}
\affiliation{International Center for Advanced Studies (ICAS) and ICIFI (CONICET), ECyT-UNSAM, Campus Miguelete, 25 de Mayo y Francia, (1650) Buenos Aires, Argentina}
\author{Gregor Srdoc}
\affil{Kotizarovci Observatory, Sarsoni 90, 51216 Viskovo, Croatia}
\author[0000-0003-2163-1437]{Chris Stockdale}
\affiliation{Hazelwood Observatory, Australia}
\author[0000-0002-7823-1090]{Paul A. Str{\o}m}
\affiliation{Department of Physics, University of Warwick, Gibbet Hill Road, Coventry CV4 7AL, UK}
\author[0000-0001-8621-6731]{Cristilyn N. Watkins}
\affiliation{Harvard-Smithsonian Center for Astrophysics, 60 Garden Street, Cambridge, MA 02138, USA}
\author[0000-0003-1452-2240]{Peter J. Wheatley}
\affiliation{Department of Physics, University of Warwick, Gibbet Hill Road, Coventry CV4 7AL, UK}
\affiliation{Centre for Exoplanets and Habitability, University of Warwick, Gibbet Hill Road, Coventry CV4 7AL, UK}
\author[0000-0001-7294-5386]{Duncan J. Wright}
\affiliation{University of Southern Queensland, West St, Darling Heights, Toowoomba, Queensland, 4350, Australia}
\author[0000-0002-4891-3517]{George Zhou}
\affiliation{University of Southern Queensland, West St, Darling Heights, Toowoomba, Queensland, 4350, Australia}
\author{Carl Ziegler}
\affiliation{Department of Physics, Engineering and Astronomy, Stephen F. Austin State University, 1936 North St, Nacogdoches, TX 75962, USA}

\author[0000-0003-2058-6662]{George Ricker}
\affiliation{Department of Physics and Kavli Institute for Astrophysics and Space Research, Massachusetts Institute of Technology, Cambridge, MA 02139, USA}
\author[0000-0002-6892-6948]{Sara Seager}
\affiliation{Department of Physics and Kavli Institute for Astrophysics and Space Research, Massachusetts Institute of Technology, Cambridge, MA 02139, USA}
\affiliation{Department of Earth, Atmospheric, and Planetary Sciences, Massachusetts Institute of Technology, Cambridge, MA 02139, USA}
\affiliation{Department of Aeronautics and Astronautics, Massachusetts Institute of Technology, Cambridge, MA 02139, USA}
\author[0000-0001-6763-6562]{Roland Vanderspek}
\affiliation{Department of Physics and Kavli Institute for Astrophysics and Space Research, Massachusetts Institute of Technology, Cambridge, MA 02139, USA}
\author{Joshua W. Winn}
\affiliation{Department of Astrophysical Sciences, Princeton University, Princeton, NJ 08544, USA}
\author[0000-0002-4715-9460]{Jon M. Jenkins}
\affiliation{NASA Ames Research Center, Moffett Field, CA 94035, USA}

\author[0000-0002-9113-7162]{Michael~Fausnaugh}
\affiliation{Department of Physics and Kavli Institute for Astrophysics and Space Research, Massachusetts Institute of Technology, Cambridge, MA 02139, USA}
\author[0000-0001-9269-8060]{Michelle Kunimoto}
\affiliation{Department of Physics and Kavli Institute for Astrophysics and Space Research, Massachusetts Institute of Technology, Cambridge, MA 02139, USA}
\author[0000-0002-4047-4724]{Hugh~P.~Osborn}
\affiliation{NCCR/Planet-S, Physikalisches Institut, Universit\"{a}t Bern, Gesellschaftsstrasse 6, 3012 Bern, Switzerland}
\author[0000-0002-8964-8377]{Samuel~N.~Quinn}
\affiliation{Harvard-Smithsonian Center for Astrophysics, 60 Garden Street, Cambridge, MA 02138, USA}
\author[0000-0002-5402-9613]{Bill Wohler}
\affiliation{SETI Institute, Mountain View, CA 94043 USA/NASA Ames Research Center, Moffett Field, CA 94035 USA}

\correspondingauthor{Emma Nabbie}
\email{Emma.Nabbie@usq.edu.au}

\begin{abstract}

The recent discoveries of Neptune-sized ultra-short period planets (USPs) challenge existing planet formation theories. It is unclear whether these residents of the Hot Neptune Desert have similar origins to smaller, rocky USPs, or if this discrete population is evidence of a different formation pathway altogether. We report the discovery of TOI-3261b, an ultra-hot Neptune with an orbital period {$P$ = {\bApproximatePeriod}} days. The host star is a $V = 13.2$ magnitude, slightly super-solar metallicity ([Fe/H] $\simeq$ 0.15), inactive K1.5 main sequence star at $d = 300$\,pc. Using data from the \textit{Transiting Exoplanet Survey Satellite} and the Las Cumbres Observatory Global Telescope, we find that TOI-3261b has a radius of {\bRadius} {\rearth}. Moreover, radial velocities from ESPRESSO and HARPS reveal a mass of {\bMass} {\mearth}, more than twice the median mass of Neptune-sized planets on longer orbits. We investigate multiple mechanisms of mass loss that can reproduce the current-day properties of TOI-3261b, simulating the evolution of the planet via tidal stripping and photoevaporation. Thermal evolution models suggest that TOI-3261b should retain an envelope potentially enriched with volatiles constituting {$\sim$}5\% of its total mass. This is the second highest envelope mass fraction among ultra-hot Neptunes discovered to date, making TOI-3261b an ideal candidate for atmospheric follow-up observations.

\end{abstract}


\section{Introduction}
\label{sec:intro}
Ultra-short period planets (USPs) are those with orbital periods $P$ $<$ 1 day and are often bombarded with extreme FUV and X-ray flux in the first hundred million years of their formation \citep{OwenWu2013}. They represent a population of highly-irradiated objects at the limits of planetary evolution. To-date, the vast majority of USP planets fall into one of two populations: super-Earths (with R $<$ 2{\rearth}) and Hot Jupiters (with $R$ $>$ 6{\rearth}). Notably absent, given the high detection efficiency of modern exoplanet surveys at these short periods, are Neptune-sized planets on $<$ 5 day orbits. This ``Hot Neptune Desert" \citep{NeptuneDesert} is interpreted as a natural consequence of planet evolution: Neptune-sized planets succumb to the extreme stellar irradiation and lose their outer layers, joining the population of small, rocky USPs \citep{OwenWu2013, LopezFortney:2013}.

However, the Transiting Exoplanet Survey Satellite (\textit{TESS}) has recently uncovered a handful of survivors, all firmly within the Hot Neptune Desert: TOI-849b \citep{Armstrong:2020}, LTT-9779b \citep{Jenkins:2020}, TOI-332b \citep{Osborn:2023}, and TOI-1853b \citep{Naponiello:2023}. All four exhibit anomalously high masses for their size, ranging from 29-73 {\mearth}. Moreover, three of these planets (TOI-849b, LTT-9779b, and TOI-332b) exhibit ultra-short orbital periods, yet somehow retain a Neptune-sized radius at this close proximity to their hosts. This reveals fingerprints of a formation pathway different from that of traditional USPs, and may indicate a different identity altogether. While rocky USPs may serve as core analogues of planets with radii up to the radius of Neptune, these Neptune-sized USPs may instead be the remnant cores of gas giants. 

We present the discovery of TOI-3261b, the fourth known Neptune-sized (3-5{\rearth}) USP with an unusually high mass ($>$ 25{\mearth}) for its size. Section \ref{sec:data} summarizes the astrometric, photometric, spectroscopic, and imaging observations of the TOI-3261 system. In Section \ref{sec:analysis}, we describe the analysis of the system parameters. This includes characterization of stellar abundances using high-resolution spectroscopy and broadband photometry, RV model selection, and the description of our global model. In Section \ref{sec:discussion}, we summarize the best-fit parameters generated by our global model and present the results of various mass loss simulations to investigate the formation and evolution of the system. We also conduct interior modeling of TOI-3261b to investigate the source of its relatively-large density. Finally, we discuss the need for future follow-up efforts in characterizing TOI-3261b to access giant planet interiors.

\section{Observations and Data analysis}
\label{sec:data}
\subsection{Astrometry \& Photometry}

TOI-3261 is a K-type star ($T_{\rm{eff}}$ = {\starTeff}\,K) with an apparent {\tess} magnitude of 12.5, as listed by the {\tess} Input Catalog \citep[TIC v8.2;][]{Stassun:2019,Paegert:2021}. 
The TIC estimates of the mass and radius of TOI-3261 are {\ticMass}{\msun} and {\ticRadius}{\rsun}, respectively. However, these parameters are less precise since TIC parameters are calculated in bulk; therefore we adopt the stellar parameters from our global modeling listed in Table \ref{tab:fit}.
Astrometry from Gaia DR3 \citep{GaiaDR3} and photometry for TOI-3261 are summarized in Table \ref{tab:star}.

\begin{deluxetable*}{lcl}
\caption{Astrometry and Photometry for TOI-3261}\label{tab:star}
\tablehead{
\colhead{Parameter} & 
\colhead{Value} & 
\colhead{Source}}
\startdata
IDs & TOI-3261 & 1\\
    & TIC 358070912 & 2\\
    & 2MASS J03105467-7331556 & 3\\
    & Gaia DR3 4640654168385886592 & 4\\
RA\,(hh:mm:ss) & 03:10:54.662 & 4\\
Dec\,(dd:mm:ss) & -73:31:55.63 & 4\\
\pmra\,(\masyr) & $50.305\pm0.012$ & 4\\
\pmdec\,(\masyr) & $34.538\pm0.011$ & 4\\
$v_{rad}$ (\kms) & $-3.14\pm1.24$ & 4\\
$\varpi$\,(mas) & $3.3278\pm0.0089$ & 4\\
$d$\,(pc) & $298.25^{+0.70}_{-0.95}$ & 5\\
SpT & K1.5$\pm$1 V & 6\\
\hline
GALEX $NUV$ & $20.482\pm0.224$ & 7\\
$B$   & 14.156 $\pm$ 0.03 & 9\\
$V$   & 13.245 $\pm$ 0.02 & 9\\
$r$ & 12.950 $\pm$ 0.05 & 9\\
$i$ & 12.739 $\pm$ 0.04 & 9\\
$T$   & 12.492 $\pm$ 0.006 & 2\\
$G$   & 13.047 $\pm$ 0.003 & 3\\
$J$   & 11.664 $\pm$ 0.022 & 3\\
$H$   & $11.271 \pm 0.025$ & 3\\
$K_s$ & $11.184 \pm 0.025$ & 3\\
\hline
$B - V$ (mag) & 0.904 $\pm$ 0.032 & 4\\
$B_p - R_p$ (mag) & $1.082 \pm 0.005$ & 4\\
$B_p - G$ (mag)   & $0.463 \pm 0.004$ & 4\\
$G - R_p$ (mag)   & $0.619 \pm 0.005$ & 4\\
E($B$-$V$) (mag) & $0.030 \pm 0.016$ & 8\\
$A_V$ (mag) & $0.096 \pm 0.051$ & 8\\
$A_G$ (mag) & $0.082 \pm 0.044$ & 8\\
$M_V$ (mag) & 5.81 $\pm$ 0.05 & 8\\
$M_G$ (mag) & 5.58 $\pm$ 0.04 & 8\\
$M_{Ks}$ (mag) & $3.795\pm0.026$ & 6\\
$U$\,(\kms) & $-79.02\pm0.65$ & 6\\
$V$\,(\kms) & $-34.83\pm0.91$ & 6\\
$W$\,(\kms) & $10.39\pm0.59$ & 6\\
$S_{tot}$\,(\kms) & $86.98\pm0.70$ & 6\\
\enddata
\tablenotetext{}{
Gaia DR3 ICRS $\alpha$ and $\delta$ positions were corrected to epoch J2000.0 via {\it Vizier}. Galactic Cartesian velocities (solar system barycentric frame): $U$ positive towards Galactic center, $V$ positive towards Galactic rotation, $W$ positive towards North Galactic Pole, and $S_{tot}$ is the total velocity with respect to the solar system barycenter \citep[following ICRS to Galactic transformations from ][]{ESA1997}.
References:
1) \citet{Kunimoto2022},
2) \citet{Stassun2019},
3) \citet{Cutri2003},
4) \citet{GaiaDR3},
5) \citet{BailerJones2021}, 
6) estimated by this work,
7) \citet{Bianchi2017}, 
8) this work (\S3), considering estimates from 
Gaia DR3 and \citet{Lallement2019},
9) \citet{Zacharias:2013}.
}
\end{deluxetable*}

\subsection{Time-Series Photometry}

\subsubsection{\textit{TESS}}
{\tess} observations of TOI-3261 (TIC 358070912) were taken from 2018 September to 2023 September. TOI-3261 was first alerted as a {\tess} Object of Interest (TOI) in Sector 13, after being initially detected by the Quick-Look Pipeline (QLP) \citep{Huang:2020b}.  Vetting through the Faint Star QLP Search \citep{Kunimoto:2022} was then conducted to identify a planet candidate around TOI-3261. The Data Validation report \citep{Huang:2020b, Huang:2020a} finds a 2.12-ppt transit signal with {$P = 0.883$} days and {$T_{dur} = 1.747$} hours. This target was selected for 2-minute cadence observations during Sectors 63, 66, 67, 68, and 69. 

The transit signature of TOI-3261 b was also detected in each transit search by the Science Process Operations Center (SPOC) in sectors 63, 66, 67, 68, and 69, and in the multi-sector search of these sectors using an adaptive, noise-compensating matched filter \citep{Jenkins:2002, Jenkins:2010, SPOC2020}. The transit signature was fitted with a limb-darkened transit model \citep{Li:2019} and subjected to a suite of diagnostic tests to evaluate the planetary hypothesis \citep{Twicken:2018}. The signature passed all the diagnostic tests including the difference image centroiding test, which for the multi-sector search constrained the location of the host star to within 0.08 {$\pm$} 2.7{\arcsec} of the transit source location. The multi-sector, multi-cadence observations of TOI-3261 are summarized in Table \ref{tab:tessobsv}.

The 30-minute and 10-minute cadence data contained in the Full Frame Images (FFIs) were detrended using a basis spline technique (see \citet{vj14, ShallueVanderburg2018, Vanderburg:2019} for a summary). When available, we use the 2-minute cadence data provided by the {\tess} SPOC pipeline \citep{Jenkins:2016}. This pipeline identifies and removes instrumental systematics via Presearch Data Conditioning Simple Aperture Photometry (PDCSAP) \citep{Stumpe:2014,Stumpe:2012,Smith:2012}. Figure \ref{fig:tessphaselc} shows the phase-folded {\tess} light curve. During light curve fitting, we perform a secondary round of detrending on both FFI and SPOC light curves using a least-squares fit of the trend to the residual, the details of which are laid out in Section \ref{subsection:global}.

\begin{deluxetable}{ccc}
\tablewidth{0pt}
\tablecaption{
  {\tess} Observation Log
  \label{tab:tessobsv}
}
\tablehead{ 
    \colhead{Sector} & \colhead{Date (UT)} & \colhead{Cadence [s]}
}
\startdata
2 & 2018 Aug 22 $-$ 2018 Sep 20 & 1800\\
6 & 2018 Dec 11 $-$ 2019 Jan 7 & 1800\\
13 & 2019 Jun 19 $-$ 2019 Jul 18 & 1800\\
27 & 2020 Jul 4 $-$ 2020 Jul 30 & 600\\
28 & 2020 Jul 30 $-$ 2020 Aug 26 & 600\\
29 & 2020 Aug 26 $-$ 2020 Sep 22 & 600\\
36 & 2021 Mar 7 $-$ 2021 Apr 2 & 600\\
63 & 2023 Mar 10 $-$ 2023 Apr 6 & 120\\
66 & 2023 Jun 2 $-$ 2023 Jul 1 & 120\\
67 & 2023 Jul 1 $-$ 2023 Jul 29 & 120\\
68 & 2023 Jul 29 $-$ 2023 Aug 25 & 120\\
69 & 2023 Aug 25 $-$ 2023 Sep 20 & 120\\
\enddata
\end{deluxetable}

\begin{figure*}[htp]
\centering
\includegraphics[width=.95\textwidth]{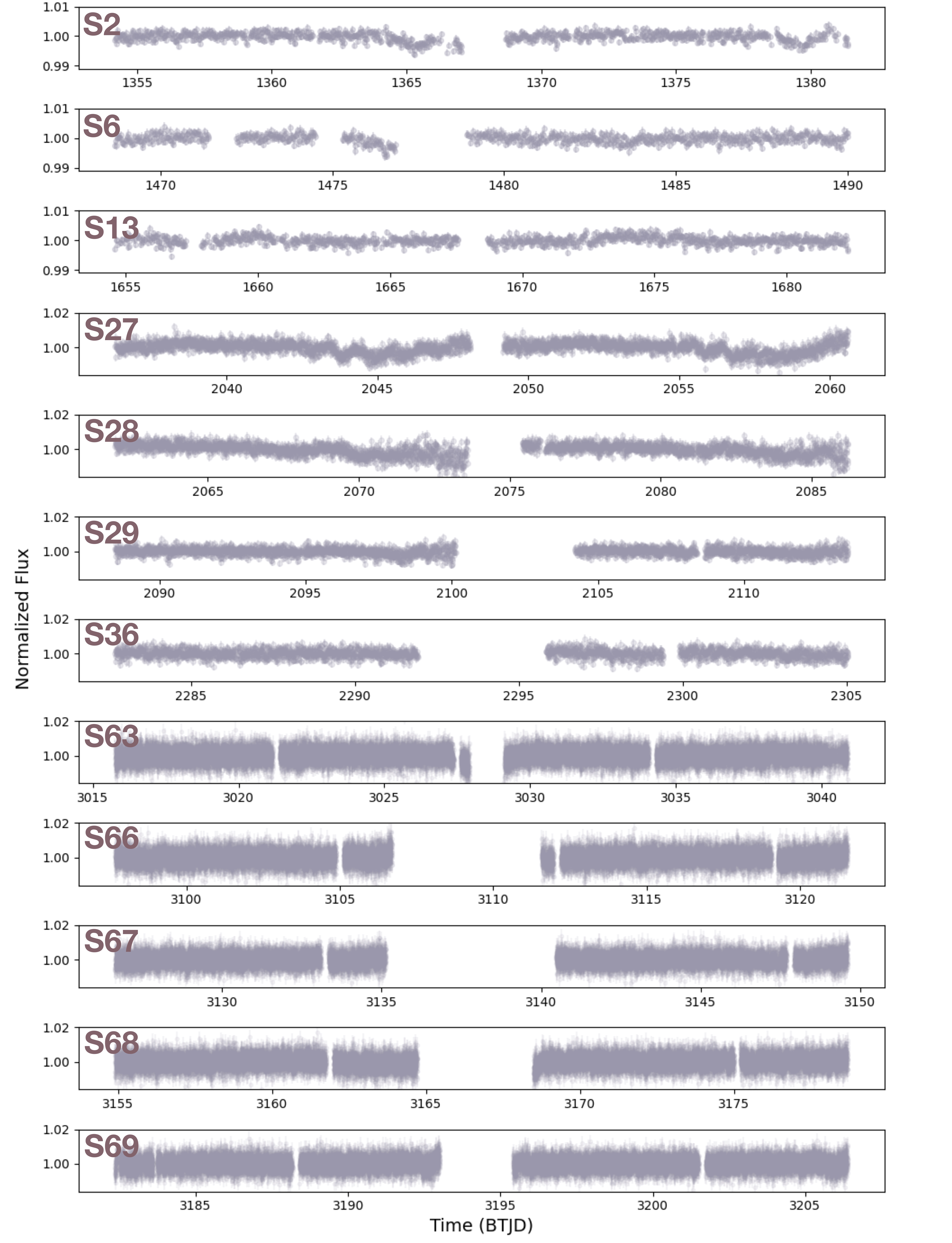}\hfill

\caption{Multi-sector, raw \textit{TESS} light curves of TOI-3261. Light curves from sectors 2, 6, and 13 are taken from FFIs with a 30-minute observing cadence. Sectors 27, 28, 29, and 36 show the 10-minute cadence FFI data. The Sector 63, 66, 67, 68, and 69 light curves depict SPOC 2-minute data.}
\label{fig:tesslc}

\end{figure*}

\subsubsection{LCOGT}
To confirm the transit signals of TOI-3261 observed by {\tess}, we make use of ground-based, seeing-limited photometry as part of the \textit{TESS} Follow-up Observing Program \citep[TFOP;][]{collins:2019}\footnote{https://tess.mit.edu/followup}. We observed three full transits of TOI-3261b in the Sloan $i'$ band using the Las Cumbres Observatory Global Telescope \citep[LCOGT;][]{Brown:2013} 1.0\,m network. The first observation was taken on UT 2023 August 4 at the 1-m node located at the South Africa Astronomical Observatory near Cape Town, South Africa (SAAO). The second full transit was observed on UT 2023 October 8 at the Siding Spring Observatory (SSO) node in New South Wales, Australia. The final full transit was captured on UT 2023 October 10 using the node located at the Cerro Tololo Inter-American Observatory (CTIO) in Chile.

Consistent among all of the 1-meter telescopes in the LCOGT network, the SINISTRO camera has a 26{\arcmin} × 26{\arcmin} field of view, with a 0.389{\arcsec} per pixel image scale. The images were calibrated by the standard LCOGT {\tt BANZAI} pipeline \citep{McCully:2018} and differential photometric data were extracted using {\tt AstroImageJ} \citep{Collins:2017}. The light curves were extracted using 2.7{\arcsec} and 5.4{\arcsec} target apertures for the SAAO and CTIO/SSO observations, respectively. There were no contamination sources present in any of these apertures. Figure \ref{fig:tessphaselc} shows the phase-folded LCO light curves, which reveal an on-target transit signal. 

\subsection{Speckle Imaging}
\label{sec:speckle}
\begin{figure}[htp]
\centering
\includegraphics[width=.5\textwidth]{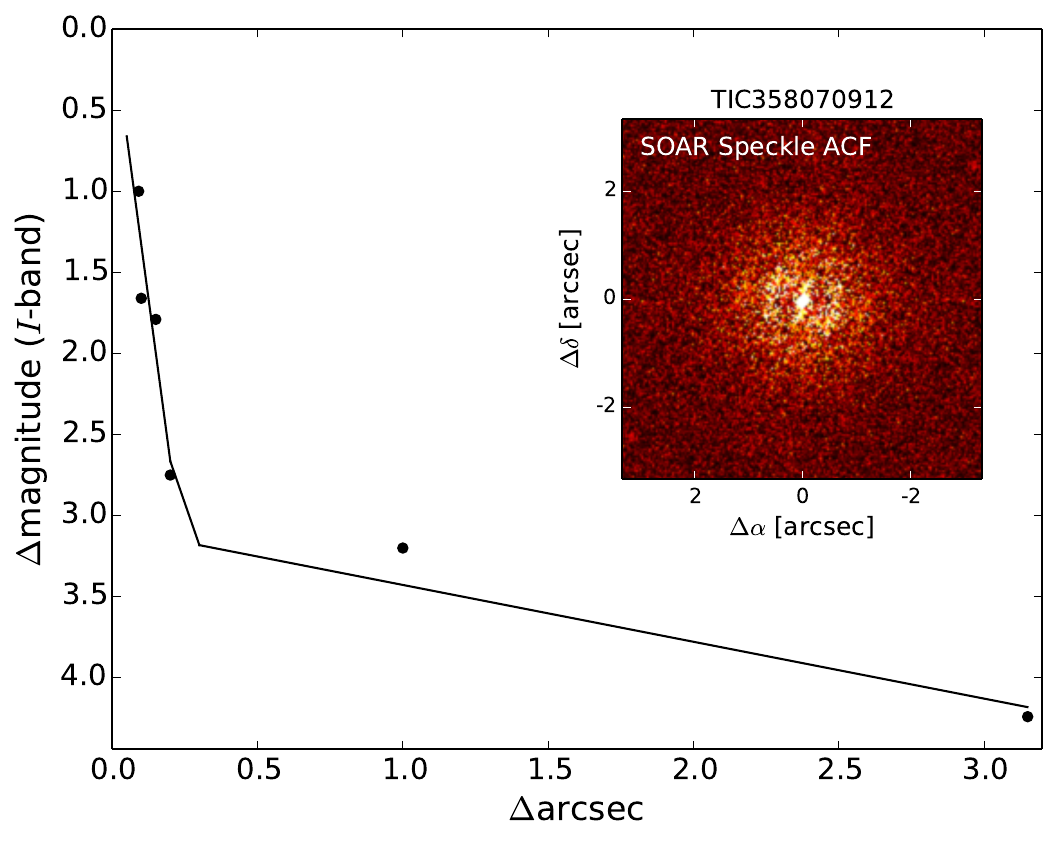}\hfill

\caption{Sensitivity curve of $I$-band delta-magnitude versus on-sky separation, with a 4{\arcsec} square image cutout inlaid. Centered on the star, the inset image shows the speckle auto-correlation function (ACF) of \target{} observed with the High-Resolution Camera on the SOAR 4.1-m telescope. It has a pixel scale of 0.01575{\arcsec} per pixel.}
\label{fig:directimage}

\end{figure}

Imaging of TOI-3261 was taken on UT 2021 July 14 with NOIRLab's High-Resolution Camera (HRCam) \citep{SoarSpeckle}. HRCam is mounted on the 4.1-m Southern Astrophysical Research Telescope (SOAR), located at the Cerro Tololo Inter-American Observatory. It has a pixel scale of 15.75 mas per pixel, with a  16{\arcsec} × 16{\arcsec} field of view and wavelength coverage from 400 to 1000 nm \citep{Tokovinin:2022}. Data reduction follows the methods described in \citet{Ziegler:2021}. These observations had an \textit{I}-band {$\Delta$}mag of 4.2 at 1{\arcsec} separation and a PSF of 0.0913{\arcsec}. The contrast curve and autocorrelation function (ACF) are shown in Figure \ref{fig:directimage}. The sensitivity curve rules out any companions with a {$\Delta$}mag of $\sim$3 within a minimum separation of $\sim$1\arcsec{}. Likewise, no companions with {$\Delta$}mag = 3 are found within 3\arcsec of TOI-3261, as the speckle ACF shows no secondary peaks at these separations.

\subsection{Time-Series Radial Velocities}

\subsubsection{HARPS}
\label{sec:harps}
We took radial velocity (RV) measurements of TOI-3261 with the High Accuracy Radial velocity Planet Searcher (HARPS) spectrograph mounted on the ESO 3.6\,m telescope at the La Silla Observatory in Chile \citep{Pepe2002}. A total of 39 spectra were obtained between UT 2023 February 3 and 2023 September 29 under the HARPS-NOMADS large programme (ID 1108.C-0697, PI: Armstrong). One observation on UT 2023 July 9 was discarded as it was forced to terminate before the exposure could be finished. The instrument (with resolving power $R = 115\,000$) was used in high-accuracy mode (HAM) with an exposure time of 2400\,s. The data were reduced using the standard offline HARPS data reduction pipeline (version 3.0.0), and a K5V template was used in a weighted cross-correlation function (CCF) to determine the RV values \citep{Baranne1996,Pepe2002}. The CCF line bisector (BIS) and full-width at half-maximum (FWHM) were measured using previously published methods \citep{Boisse2011}, and no correlations were found between the RVs and FWHM, BIS, or the pipeline-extracted S-index. The exposures had a median radial velocity error of 13.48 m s$^{-1}$, and an RMS of 13.24 m s$^{-1}$ after the planetary signal was removed. These uncertainties reflect only the contribution from photon noise. The RV measurements can be found in Table \ref{tab:rv}.

\subsubsection{ESPRESSO}
As part of ESO Program 111.24W6 (PI: C. Huang), we obtained 15 observations of TOI-3261 with the ESPRESSO spectrograph \citep{Pepe2021}. Time-series radial velocities were taken from UT 2023 September 19 to 2023 September 22 in ESPRESSO's high-resolution 1UT mode ($R = 140\,000$), with $2\times1$ binning and slow read-out. All measurements were taken with UT2 and exposed to a set SNR of 30 at 550nm (Order 102). The exposures had a median radial velocity error of 2.07 m s$^{-1}$, and an RMS of 3.41 m s$^{-1}$ after removing the planet's signal. These uncertainties only reflect the contribution from photon noise. The combined HARPS + ESPRESSO datasets are shown in Figure \ref{fig:rv}, as well as summarized in Table \ref{tab:rv}.

The data were processed using Version 3.1.0 of the ESPRESSO Data Reduction Software \citep[DRS;][]{Pepe2021} \footnote{https://www.eso.org/sci/software/pipelines/espresso/} which produces 1-D, order-by-order, wavelength calibrated spectra, RV measurements, and CCF metrics such as the BIS and FWHM. To calculate RVs, a K2V mask was used to produce order-by-order CCFs of the stellar spectrum, following the procedure outlined in \citet{Baranne1996} and \citet{Pepe:2000}.

\section{Analysis}
\label{sec:analysis}
We obtain stellar parameter estimates and constrain the abundances of individual elements through combined spectroscopic and photometric analyses. To demonstrate consistent results and ensure the accuracy of our stellar parameters, we make use of two separate packages to characterize TOI-3261. 

\subsection{Extinction Estimate}
The parallax from Gaia DR3 \citep{Gaia,GaiaDR3} was used to constrain the distance. Following the recommendation by \citet{DR3ParallaxCorrection}, we corrected the Gaia parallax to mitigate the systematic offset based on stellar G-band magnitude, position, and color. These priors are used during the simultaneous fitting of stellar parameters in our global model (see Section \ref{subsection:global} for a detailed summary).

The reddening for the star is small but not zero. The STILISM 3D reddening maps \citep{Lallement2019} predict \EBV\ = $0.038^{+0.006}_{-0.016}$ mag, which means most of the Galactic reddening in this direction \citep[E($B$-$V$)$_{max}$ = 0.052;][]{Schlegel1998} is foreground to TOI-3261. The TIC adopts a reddening value informed by the STILISM maps: \EBV\ = $0.0375\pm0.0102$ mag \citep{Stassun2019}.
However, Gaia DR3 provides lower estimates
$A_0$ = $0.0486^{+0.0070}_{-0.0071}$ \citep[$\simeq$ $A_V$ $\simeq$ 3.2\,\EBV for
early K dwarfs;][]{McCall2004}. Using the +1$\sigma$ STILISM value as an upper bound
and the -1$\sigma$ Gaia DR3 value as a lower bound, our estimate of the reddening is \EBV\, = $0.030\pm0.016$ mag or extinction $A_V$ = $0.096\pm0.051$ mag or $A_G$ = $0.082\pm0.044$ mag
\citep[adopting $A_G$/\EBV\, = 2.74 from ][]{Qin2023}. 
Combining these extinction estimates with Gaia DR3 photometry and parallaxes, we estimate the star's absolute magnitudes to be 
$M_V$ = $5.81\pm0.05$ and $M_G$ = $5.58\pm0.04$, consistent with a main sequence K0V star. 

\begin{figure}[htp]
\centering
\includegraphics[width=.5\textwidth]{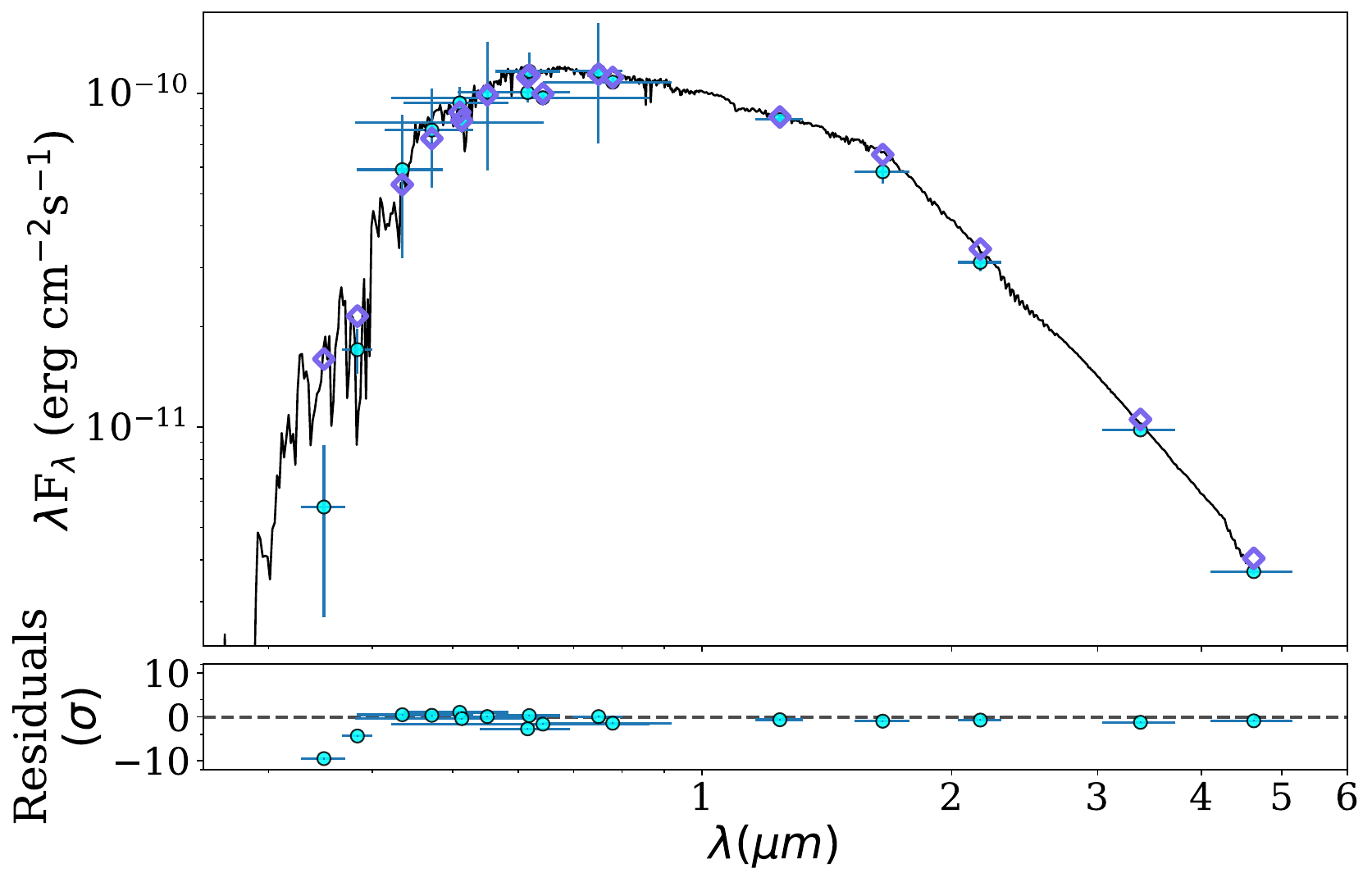}\hfill

\caption{{Stellar spectral energy distribution with an ATLAS9 model 
($T_{eff}$ = 5000 K, [Fe/H] = 0 dex, log($g$) = 4 dex) overlaid. Individual points represent photometric data from the \textit{Gaia} DR2 Crossmatch Catalog. Horizontal errors denote the bandpass widths.}}
\label{fig:SED}

\end{figure}

\subsection{Individual Stellar Abundances}
\label{sec:abundances}

\subsubsection{ARES}
The stellar spectroscopic parameters ($T_{\mathrm{eff}}$, $\log g$, microturbulence, [Fe/H]) were estimated using the ARES+MOOG methodology \citep[][]{Sousa-21, Sousa-14, Santos-13}. The equivalent widths (EWs) were consistently measured using the ARES code\footnote{The latest version, ARES v2, can be downloaded at \url{https://github.com/sousasag/ARES}} \citep{Sousa-07, Sousa-15}. We used the list of iron lines appropriate for stars cooler than 5200 K, which was presented in \citet[][]{Tsantaki-2013}. We adopt the values from our analysis of a combined ESPRESSO spectrum due to the higher signal-to-noise data. Nevertheless, find that of the HARPS combined spectra gives consistent but less precise results. To converge for the best set of spectroscopic parameters for each spectrum we use a minimization process to find the ionization and excitation equilibrium. This process makes use of a grid of Kurucz model atmospheres \citep{KuruczModel} and the latest version of the radiative transfer code MOOG \citep{Sneden-73}. We also derived a more accurate trigonometric surface gravity using recent GAIA data following the same procedure as described in \citet[][]{Sousa-21} which provided a consistent value when compared with the spectroscopic surface gravity. 

Using the aforementioned stellar atmospheric parameters (we considered spectroscopic surface gravity), we determined the abundances of Mg, Si, and Ni  following the classical curve-of-growth analysis method described in \citet[e.g.][]{Adibekyan-12, Adibekyan-15}. Similar to the stellar parameter determination, we used ARES to measure the EWs of the spectral lines of these elements, and used a grid of Kurucz model atmospheres along with the radiative transfer code MOOG to convert the EWs into abundances, assuming local thermodynamic equilibrium. Although the EWs of the spectral lines were automatically measured with ARES, for Mg which has only three lines available we performed careful visual inspection of the EW measurements. The results from ARES are located in Table \ref{tab:abundances}.

\begin{table}
\centering
\caption{ARES Fit Results for TOI-3261}
\label{tab:abundances}
\begin{tabular}{lccl}
\hline
Parameter [unit] & HARPS & ESPRESSO\tablenotemark{a} & Source\\
\hline
HARPS\\
\hline
$T_\mathrm{eff}$ [K]  & $5077  \pm 82$ & $5065  \pm 72$               & This work\\
$\log g$ [dex]        & $4.28  \pm 0.15$  & $4.19  \pm 0.13$             & This work (Spec)\\
$\log g$ [dex]        & $4.48  \pm 0.04$  & $4.48  \pm 0.05$            & This work (Gaia)\\
$[\mathrm{Fe/H}]$ [dex]  & $0.15  \pm 0.05$  & $0.14  \pm 0.02$            & This work\\
$[\mathrm{Mg/H}]$ [dex]  & $0.13 \pm 0.08$   & $0.17 \pm 0.05$           & This work\\
$[\mathrm{Si/H}]$ [dex]  & $0.24 \pm 0.06$   & $0.22 \pm 0.09$           & This work\\
$[\mathrm{Ni/H}]$ [dex]  & $0.20 \pm 0.06$  & $0.17 \pm 0.06$            & This work\\
$v_\mathrm{mic}$ [km~s$^{-1}$] & $0.61  \pm 0.15$ & $0.57  \pm 0.10$    & This work\\ 
\hline
\end{tabular}
\tablenotetext{a}{We adopt these values as priors for stellar fitting.}

\end{table}

\subsubsection{SPECIES + ARIADNE}
As a check on the results of the fit from ARES, we use a combined {\tt SPECIES+ARIADNE} methodology as a secondary analysis of stellar parameters. From the merged, 1D HARPS spectrum, we obtained individual stellar abundances and bulk parameters using the {\tt SPECIES} package \citep{species}. This package makes use of the MOOG algorithm and a custom line fitting routine to help better deal with line blends called EWComputation (see \citet{Soto:2021}), along with fitting to ATLAS9 stellar atmosphere models. This generates stellar parameters that are then used as priors for a Bayesian Model Averaging (BMA) approach with {\tt ARIADNE} \citep{ariadne}.

Through {\tt ARIADNE}, we used the star's broadband photometry to construct a spectral energy distribution (SED). The SED is shown in Figure \ref{fig:SED}. The following photometric bands were included: SkyMapper $u$, $v$, and $r$ \citep{Onken:2019}; Gaia DR2 {$B_{P}$}, $G$, and {$R_{P}$} \citep{GaiaDR3}; 2MASS $J$, $H$, and {$K_{s}$} \citep{Skrutskie2006}; SDSS $g$, $r$, and $i$ \citep{York:2000}; Johnson $B$ and $V$ \citep{Johnson:1953};and WISE $W1$ and $W2$ \citep{Wright2010}. The SED was then fit to the following stellar models: PHOENIXv2 \citep{Husser2013}, BTSettl-AGSS2009 \citep{Allard2011}, Kurucz \citep{KuruczModel}, and Castelli \& Kurucz (ATLAS9) \citep{Castelli2004}. The BMA method calculates the weighted average of the {\teff}, radius, {\logg}, and {\feh} estimates generated by each model. This weighted average accounts for many of the biases that might be present across different atmospheric models. 

The results from the SPECIES + ARIADNE fit are shown in Table \ref{tab:species}. The estimated stellar abundances are consistent with those from ARES, demonstrating the robustness of the results from spectroscopic and photometric analysis. The stellar parameters are well-constrained regardless of which methodology was used, which in turn lends greater confidence to the robustness of our planet parameters. In our analysis, we then adopt the ARES parameters estimated from the ESPRESSO spectra. We adopt the \logg from ESPRESSO in Table \ref{tab:abundances}, while the {\teff} and {\feh} values from ESPRESSO are used as priors in our global model (Section \ref{subsection:global}).

\begin{table}
\centering
\caption{SPECIES+ARIADNE Fit Results for TOI-3261}
\label{tab:species}
\begin{tabular}{lcc}
\hline
Parameter & Value &  \\

\hline
ARIADNE\\
\hline
$T_\mathrm{eff}$ [K]     & $4995  \pm 47$     & \\
$\log g$ [dex]           & $4.15  \pm 0.22$   & \\
$[\mathrm{Fe/H}]$ [dex]  & $0.112  \pm 0.075$   & \\
Radius [\rsun] & $0.906 \pm 0.024$ & \\
\hline

SPECIES\\
\hline
$T_\mathrm{eff}$ [K]     & $5063  \pm 58$     & \\
$\log g$ [dex]           & $4.13  \pm 0.13$   & \\
$[\mathrm{Fe/H}]$ [dex]  & $0.11  \pm 0.04$   & \\
$vsini$ [km s$^{-1}$]    & $2.83 \pm 0.21$ & \\
$v_{t}$ [km s$^{-1}$]    & $0.33 \pm 0.23$ & \\
$v_{mac}$ [km s$^{-1}$]  & $1.51 \pm 0.21$ & \\
$\#$ Fe I Lines          & 129 & \\
$\#$ Fe II Lines         & 18  & \\
\hline

Element & Value & $\#$ Lines\\
\hline
$[\mathrm{Al/H}]$   & $0.01 \pm 0.12$    & 3 \\
$[\mathrm{Ca/H}]$   & $0.15 \pm 0.1$      & 4 \\
$[\mathrm{Cr/H}]$  & $0.14 \pm 0.06$    & 13 \\
$[\mathrm{Cu/H}]$  & $0.54 \pm 0.12$    & 3 \\
$[\mathrm{FeI/H}]$  & $0.18 \pm 0.055$    & 13 \\
$[\mathrm{FeII/H}]$ & $-0.06 \pm 0.12$    & 3 \\
$[\mathrm{Mg/H}]$   & $0.15 \pm 0.12$    & 3 \\
$[\mathrm{Mn/H}]$   & $0.36 \pm 0.10$    & 4 \\
$[\mathrm{Na/H}]$   & $0.18 \pm 0.12$    & 3 \\
$[\mathrm{Ni/H}]$   & $0.22 \pm 0.08$    & 6 \\
$[\mathrm{Si/H}]$   & $0.08 \pm 0.08$    & 6 \\
$[\mathrm{TiI/H}]$   & $0.14 \pm 0.06$    & 10 \\
$[\mathrm{TiII/H}]$  & $0.25 \pm 0.12$    & 3 \\

\hline

\end{tabular}

\end{table}

\subsection{Selecting an RV model \label{sec:rvmodel}}

Before implementing the combined transits + RV + SED fit, we make use of the \texttt{RadVel} software package \citep{Fulton2018} to fit just the HARPS and ESPRESSO RV data for TOI-3261 b and decide what basic RV model to adopt. We test five different models for the system, all of which include a single planet on a circular orbit with a period initialized at the best-fit TESS transit period of TOI-3261 b. The first model incorporates only planet b, the next three models also include a linear and/or quadratic trend, and the fifth model combines the planet with a 1D, quasi-periodic, Gaussian Process (GP) trained on the RV data  (Table \ref{tab:rvmodels}).

\begin{figure*}[htp]
\centering
\includegraphics[width=.8\textwidth]{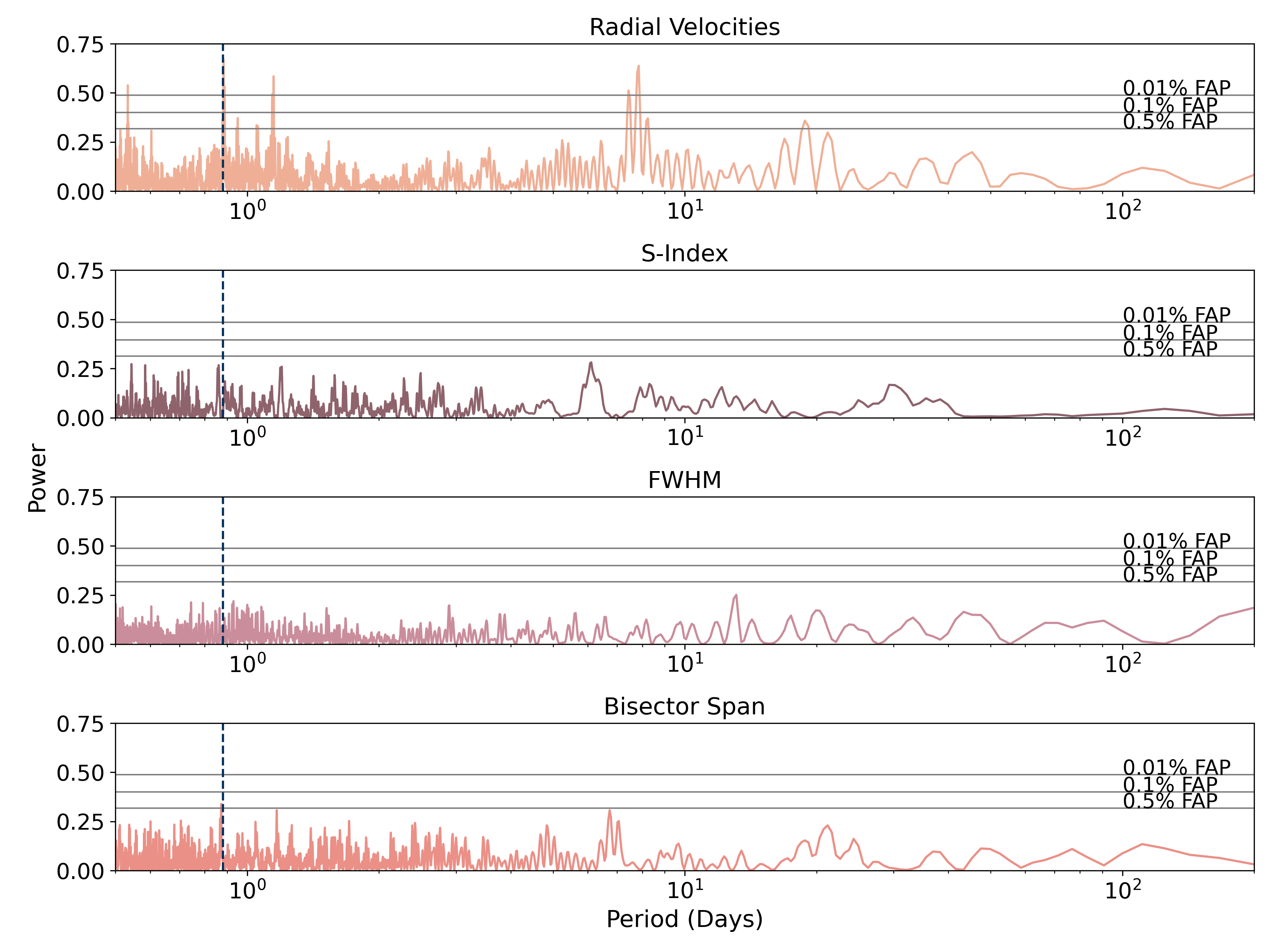}\hfill

\caption{Lomb-Scargle periodograms derived from HARPS observations. The vertical dashed line denotes the period of TOI-3261b, while the horizontal lines represent computed 0.5\%, 0.1\%, and 0.01\% False Alarm Probabilities. The topmost panel shows a peak at the planetary period, but the periodograms of the activity indicators (lower three panels) show no significant peaks near the 0.88-day mark. The peak at $\sim$8 days in the top panel is not astrophysical, as it disappears once the planet signal is removed.}
\label{fig:periodograms}

\end{figure*}

Neither the TESS light curves (Figure \ref{fig:tesslc}) nor the radial velocities and activity indicators derived from the HARPS spectra (Figure \ref{fig:periodograms}) exhibit the quasi-periodic sinusoidal behavior generally induced by rotationally modulated surface features such as star spots or plages \citep{SaarDonahue1997}. Without such features to generate obvious variations in these time series data, constraining the rotational period of the star proves challenging and we are only able to produce a rough estimate of the star's \Prot\ from its rotational velocity. Combining the \vsini\, value derived from the ESPRESSO data with \texttt{SPECIES} (2.83 $\pm$ 0.21 \kms) with the best-fit stellar radius from \texttt{ARIADNE} (0.906 $\pm$ 0.024 \rsun) suggests an upper limit for \Prot\, at approximately 16 days.

Given the lack of rotational modulation in the light curves and RV data products, we don't expect a strong rotational activity component to be present in the RV data. When adding the GP to our fifth possible RV model, we initialize it with a broad uniform prior on the period (10 - 20 days) and a broad Gaussian prior on the spot lifetime ($\mu$ = 45 days, $\sigma$= 16 days).

\texttt{Radvel} computes the Bayesian Information Criterion ($\rm{BIC} = -2\ \rm{ln}\ \mathcal{L}_{\rm{max}}\ + \it{n}\ \rm{ln}\ \it{N}$, where $\mathcal{L}_{\rm{max}}$ is the maximum likelihood, $\emph{n}$ is the number of model free parameters, and $\emph{N}$ is the number of data points) value for each model as applied to our data, which we compile in Table \ref{tab:rvmodels}. Models with lower BIC values are preferred \citep{KassRaftery1995}, and we find that the simplest model of a single, circular planet with a period initialized on that of TOI 3261 b has the lowest BIC value. The $\Delta$BIC between this model and those incorporating a linear or quadratic trend is not significant, and given the lack of Bayesian evidence for a more complicated model we opt to use the single circular planet model in the global fit. We do not consider an eccentric planet model, as tidal circularization of a planet with TOI-3261b's mass and location would take $\sim$10{$^{7}$} years (Eq. 5 of \citet{Papaloizou:2010}) assuming TOI-3261b has a tidal quality factor $Q$ and Love number {$k_2$} similar to those of Neptune. Since this is two orders of magnitude below the age of the system, tidal circularization should have been completed. Additionally, we do not see any evidence of trends that could point to additional companions, as the feature in the radial velocity periodogram at $\sim$8 days (top panel of Figure \ref{fig:periodograms}) does not remain after TOI-3261b's signal is removed.

\begin{deluxetable}{lcc}
\tablewidth{0pt}
\tablecaption{
  RV Model Selection
  \label{tab:rvmodels}
}
\tablehead{ 
    \colhead{Model} & \colhead{BIC}
}
\startdata
1 cp\tablenotemark{a} & 396.66 \\
1 cp + linear & 399.08 \\
1 cp + quadratic\tablenotemark{b} & 399.42\\
1 cp + linear + quadratic & 402.95 \\
1 cp + GP & 412.7 \\
\enddata
\tablenotetext{a}{``cp" denotes a planet on a circular orbit.}
\tablenotetext{b}{Only a second-order trend is included.}
\end{deluxetable}

\subsection{Global Modeling of System Parameters}
\label{subsection:global}

To fully characterize the system architecture, we use a Markov Chain Monte Carlo (MCMC) method to jointly fit transit, isochrone, and radial velocity data. This is performed using the {\tt emcee} package \citep{ForemanMackey:2012}. By taking into account both spectroscopic and photometric data, we are able to obtain a more comprehensive view of the system. The global likelihood function then consists of transit, isochrone, and radial velocity components, each with their own individual likelihoods.

For the transit component, {\tess} and LCO light curves are fit with a {\tt batman} transit model \citep{batman}. For the {\tess} observations, the model includes a supersampling factor corresponding to each observing cadence. In each step of the MCMC, we simultaneously detrend the light curves with a least-squares method independently for each instrument. We co-trend with the PSF width for LCO data, and we detrend with a third-order polynomial for the \textit{TESS} data.

Data from each node of the LCOGT network is fit separately to account for variable observing conditions and instrument systematics. Treating each node as a separate instrument, detrending is conducted individually to mitigate any telescope-specific artifacts.

Limb darkening coefficients for each instrument are also left as free parameters, with each instrument having its own set of limb darkening parameters. For the ground-based transits, we fit for one set of limb darkening parameters that are shared across all three LCO transits, since they were all captured in the \textit{i'} band.

Aside from estimating planet parameters from the light curve, stellar parameters are simultaneously modeled using MIST isochrones \citep{MISTModels}, compiled with the {\tt isochrones} package \citep{Isochrones}. This affords greater precision in both stellar and planet parameter estimates, and allows us to investigate the star's approximate mass and age. We then fit the interpolated stellar photometry to the star's broadband photometric data.

We use RadVel \citep{radvel} to fit the combined HARPS + ESPRESSO data set, applying separate RV offset and RV jitter terms to each instrument. These individual offset ({$\gamma$}) and jitter ({$\sigma$}) terms for each instrument were included as free parameters, while the radial velocity semi-amplitude was a shared parameter across both instruments. We did not model any linear or quadratic trends, assuming a one-planet system with a circular orbit (see Section \ref{sec:rvmodel} for more details).

Subsequently, the total log likelihood was then the sum of the transit, isochrone, and RV log likelihoods. In total, the following free parameters were included in our global fit: [{$p$, $t_{0}$, $q_1$, $q_2$, ${\frac{R_p}{\rstar}}$, $b$, $K$, {\mstar}, {\rstar}, {\teff}, {${\feh}$}, {$\gamma$}, {$\sigma$}, age, error scale}]. Parallax, distance, and extinction remained fixed. Limb darkening coefficients were re-parametrized as {$q_1$} and {$q_2$} following the methods in \citet{Kipping:2013}. Gaussian priors on {\teff} and {\feh} were implemented using estimates from the ARES fit of ESPRESSO data, while the period and transit epochs had Gaussian priors using the values and errors from ExoFOP. Uniform priors on stellar radius and mass were implemented based on the estimates from ESPRESSO. The priors for each free parameter are summarized in Table \ref{tab:fit}. 

\subsection{Confirmation with Ground-Based Facilities}
We confirm TOI-3261b as a planet through multiple, independent detection methods. From ground-based observatories, we find on-target transit signals, reproducing the signals first seen by {\tess}. These cannot be attributed to any stellar companions, as this case is ruled out by SOAR speckle imaging to the sensitivity limits described in Section \ref{sec:speckle}. Both HARPS and ESPRESSO radial velocities reveal a planet-mass object, with no BIS,FWHM, or S-index variability at the periodicity of the planetary signal, further excluding any astrophysical false positive scenarios.

\section{Results and Discussion}
\label{sec:discussion}
\subsection{Best-Fit System Parameters}
\begin{figure}[htp]
\centering
\includegraphics[width=.48\textwidth]{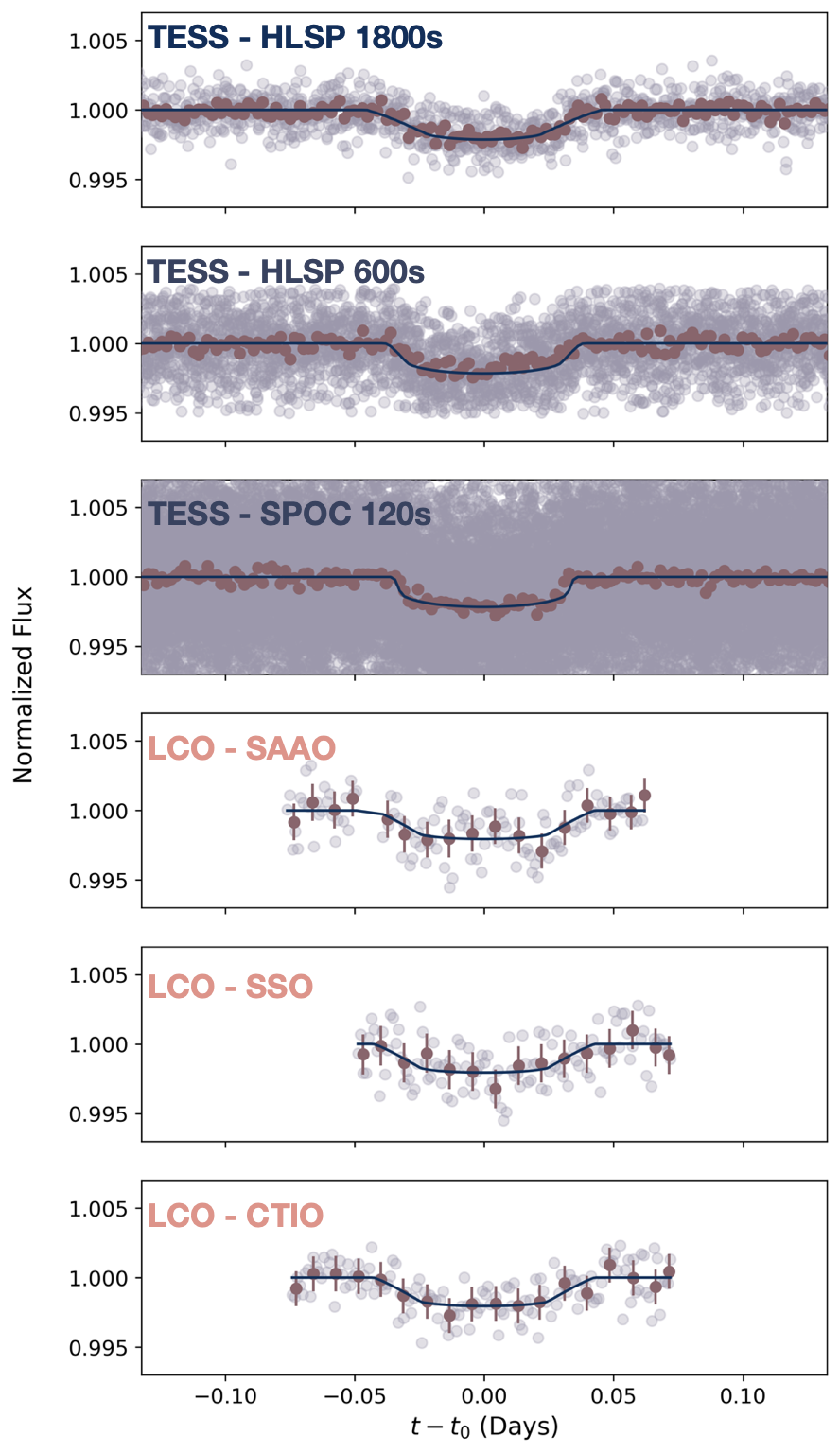}\hfill

\caption{{\textit{Top three panels:}} Phase-folded {\tess} FFI/HLSP and SPOC light curves of \target{}, with binned data in mauve and a best-fit {\tt batman} transit model overlaid. Data taken at different cadences are displayed separately, as the observing cadences require different supersampling of the light curve to account for the change in transit shape. {\textit{Bottom three panels:}} Phase-folded LCO light curves, with binned errorbars included. These panels employ the same color scheme as the top three.}
\label{fig:tessphaselc}

\end{figure}

\begin{figure}[htp]
\centering
\includegraphics[width=.5\textwidth]{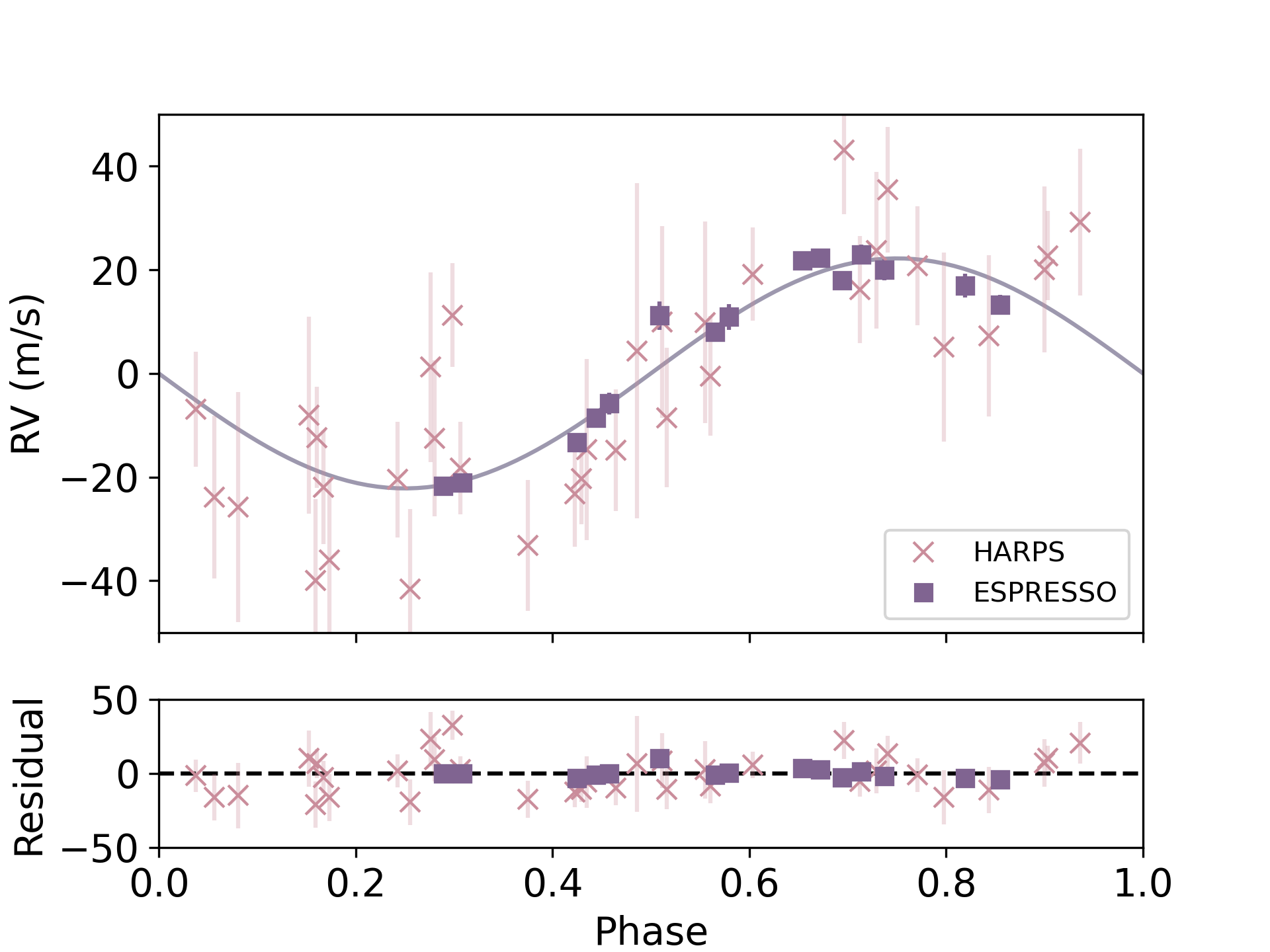}\hfill

\caption{{Phase-folded HARPS + ESPRESSO radial velocities of TOI-3261, with a best-fit {\tt RadVel} model overlaid and residuals in the bottom panel. The pink crosses represent the HARPS data points, and the purple squares denote ESPRESSO data.}}
\label{fig:rv}

\end{figure}

\begin{figure}[htp]
\centering
\includegraphics[width=.5\textwidth]{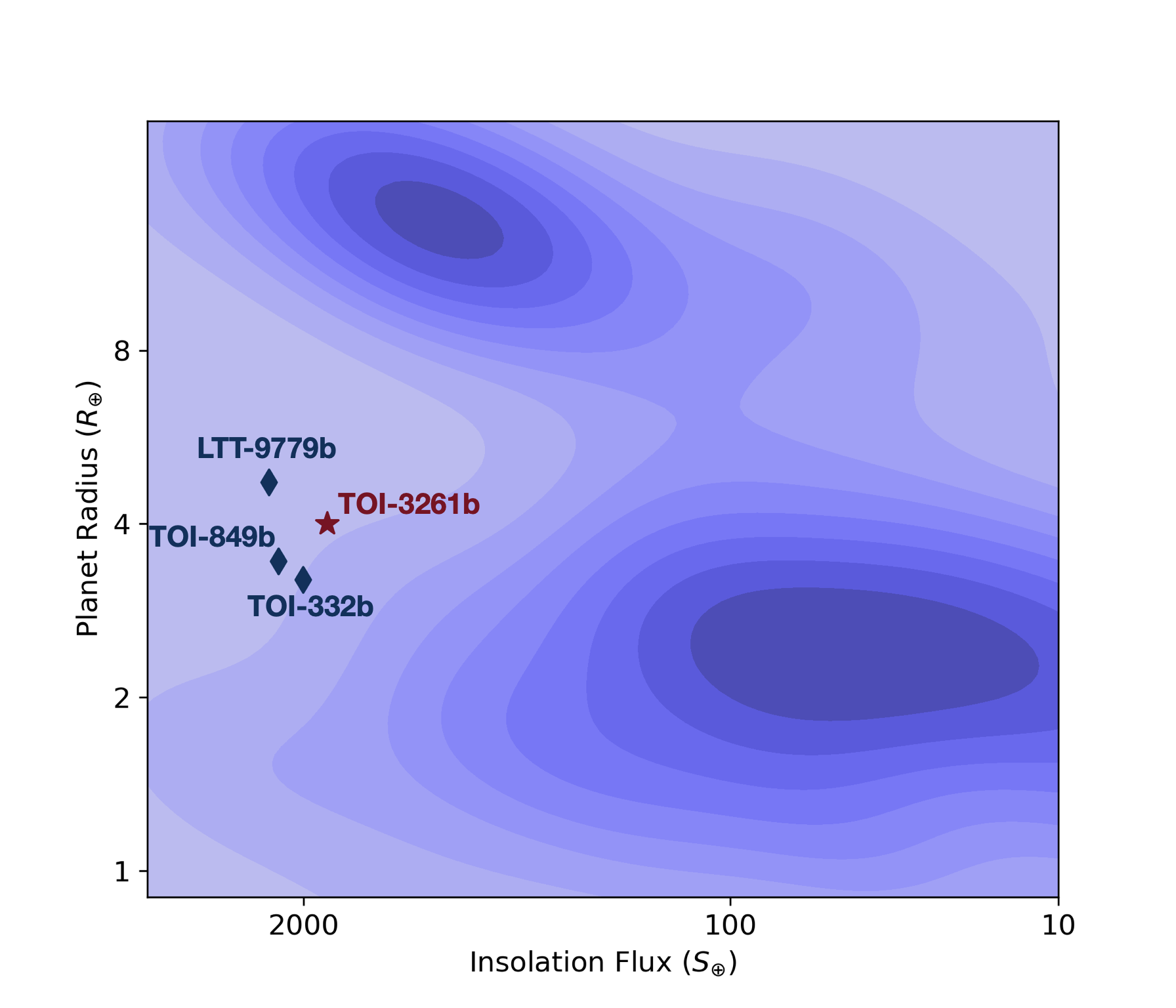}\hfill

\caption{TOI-3261b in context with the published, confirmed exoplanet population as marked by the NASA Exoplanet Archive, as of 2024 February 4. The density map shows the distribution of confirmed exoplanets in radius-insolation space. The other Neptune-sized USPs are denoted in dark blue, with TOI-3261b represented in red. TOI-3261b sits squarely within the Hot Neptune Desert, shown by the dearth of Neptune-sized exoplanets at high irradiations.}
\label{fig:population}

\end{figure}

The median results from the joint transit, SED, and RV modeling are summarized in Table \ref{tab:fit}. The best-fit models to \textit{TESS} and LCO photometry are shown in Figure \ref{fig:tessphaselc}, while the best-fit RV model is shown in Figure \ref{fig:rv}. Our global fit confirms that TOI-3261b is the fourth-known Neptune-sized (3-5 {\rearth}) USP (Figure \ref{fig:population}), with an orbital period of {\bApproximatePeriod} days and a radius of {\bRadius} {\rearth}. Moreover, TOI-3261b has a mass of {\bMass} {\mearth}, leading to a bulk density of {\bDensity} {g cm$^{-3}$} and placing it within the population of overly-dense USP Neptunes (whose density is greater than Neptune's density of 1.64 g cm$^{-3}$). Its large radius suggests that TOI-3261b retains a gaseous envelope, with a core potentially enriched with water (Figure \ref{fig:massradius}). As TOI-3261b is only the fourth-discovered object of its kind, it presents another valuable opportunity to test formation theories of these unusual planets.

We also estimate the age of TOI-3261 using its best-fit mass and metallicity to correlate the activity indices from ESPRESSO spectra. We extracted CaII index values from ESPRESSO spectra using ACTIN\footnote{\url{https://actin2.readthedocs.io/en/latest/index.html}} \citep{GomesdaSilva:2018,GomesdaSilva:2021}, and converted them to the log$(R'_{HK})$ index following the method described in \citet{GomesdaSilva:2021}. We find a median log$(R'_{HK})$ Ca II index of -5.16 {$\pm$} 0.05 dex. Using the age-mass-metallicity-activity relation from \citet{Lorenzo-Oliveira:2016}, this gives an age of 6.5 {$\pm$} 2.1 Gyr for TOI-3261. This methodology accounts for the biases in mass and metallicity to readjust the activity-age relation, which provides an age estimate more correlated to asteroseismic ages.

One result of note is TOI-3261's metallicity ([Fe/H] = {\starfeh} dex). Like the other USP Neptunes, TOI-3261b is hosted by a metal-rich star. This preference of ultrahot Neptunes to form around high-metallicity stars was established by \citet{Dai:2021}, which aligns with Hot Jupiters' predisposition to form around metal-rich stars \citep{FischerValenti2005}. On the other hand, terrestrial USPs do not share the same preference to form around higher-metallicity hosts \citep{Winn:2017}. This lends credence to the notion that ultrahot Neptunes are a distinct population from smaller, terrestrial USPs, and may be descended instead from Hot Jupiters. 

\subsection{Planet Composition and Photo-evaporative Modeling}
We model the photoevaporation history of TOI-3261b to understand its potential contribution to the planet’s overall mass loss. Using the Python code {\tt photoevolver} \citep{Fernandez-Fernandez:2023}, we solve for the internal structure of the planet and model the evolution of its atmosphere due to photoevaporation. We approximate TOI-3261b’s interior with a two-layered model, consisting of a solid core and a pure H/He envelope. In this framework, all of the planet’s metals lie within the core. Under these assumptions, we find that the planet’s current-day properties imply an envelope mass fraction of {5.1$\pm$2.6\%}. This is the second highest envelope mass fraction among all Neptune-sized USPs known to date. Our model also uses the calculations from \citet{LopezFortney:2014} to estimate an envelope thickness of 1.3$\pm$0.3 {\rearth}. 

\begin{figure*}[htp]
\centering
\includegraphics[width=.8\textwidth]{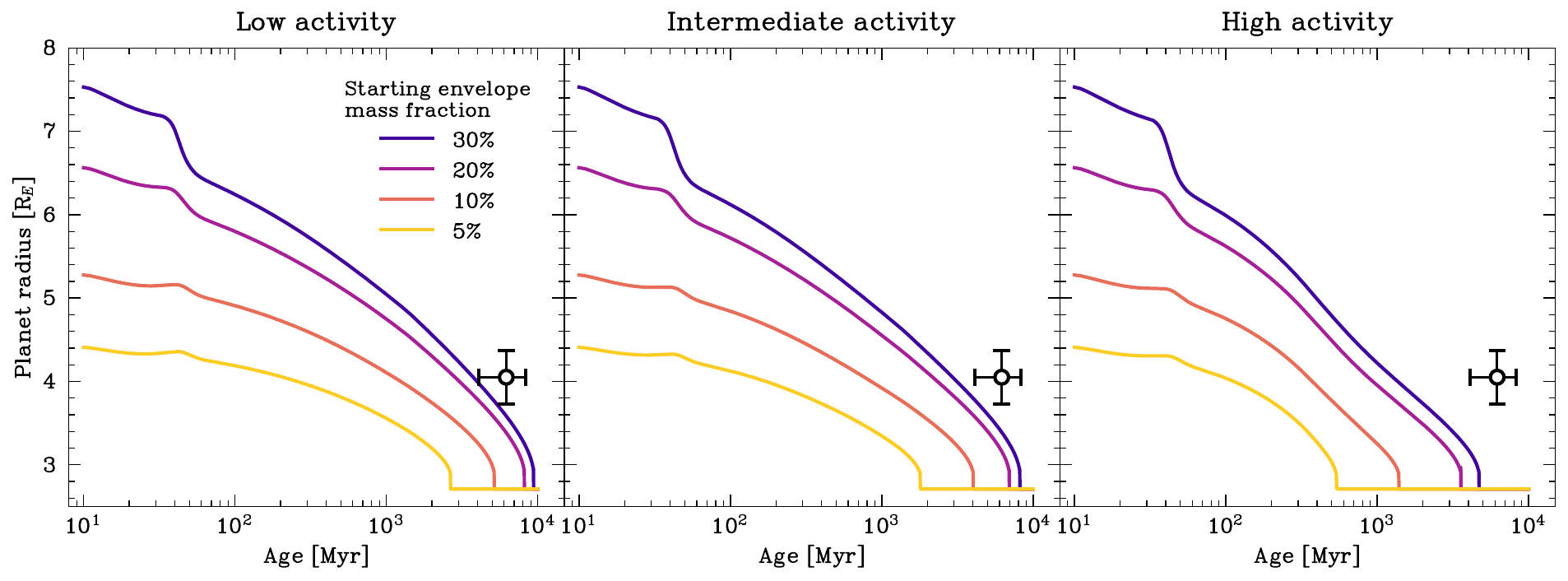}\hfill

\caption{Evolution of the radius of TOI-3261b as a function of varying initial envelope mass fractions, with a starting age of 10 Myr. These simulations assume a pure H/He envelope. The left, middle and right panels show the planet's radius evolution while experiencing low, intermediate, and high levels of stellar irradiation respectively. The empty circle denotes the planet's current radius and age, with the associated errorbars for both parameters.}
\label{fig:evo}

\end{figure*}

Stellar XUV flux was estimated through stellar evolution grids from \citet{Johnstone:2021}, assuming a stellar age of 6.5 {$\pm$} 2.1 Gyr. We vary our assumptions of the stellar irradiation flux (ranging from low, intermediate, and high) to investigate its effect on the mass loss of TOI-3261b. These values are taken from the 2-{$\sigma$} lower, middle, and upper bounds of the stellar XUV flux estimated by the \citet{Johnstone:2021} stellar evolution tracks. On this timescale, we simulate the photoevaporation history of TOI-3261b to infer its original size. These simulations adopt the non-energy-limited mass loss formulation from \citet{Kubyshkina:2018}, which accounts for other effects of evaporation such as recombination. Initializing the simulations at 10 Myr, we track the evolution of the planet's radius with various initial envelope mass fractions (Figure \ref{fig:evo}), constructing the envelope structure using models from \citet{ChenRogers2016}. Regardless of irradiation flux or starting envelope mass fraction, these models find high rates of escape if the envelope was solely composed of light elements like H/He, with none of the scenarios allowing the planet to retain its envelope. Therefore, the atmosphere of TOI-3261b cannot be made of pure H/He if the envelope were to survive to present day.

\begin{figure}[htp]
\centering
\includegraphics[width=.5\textwidth]{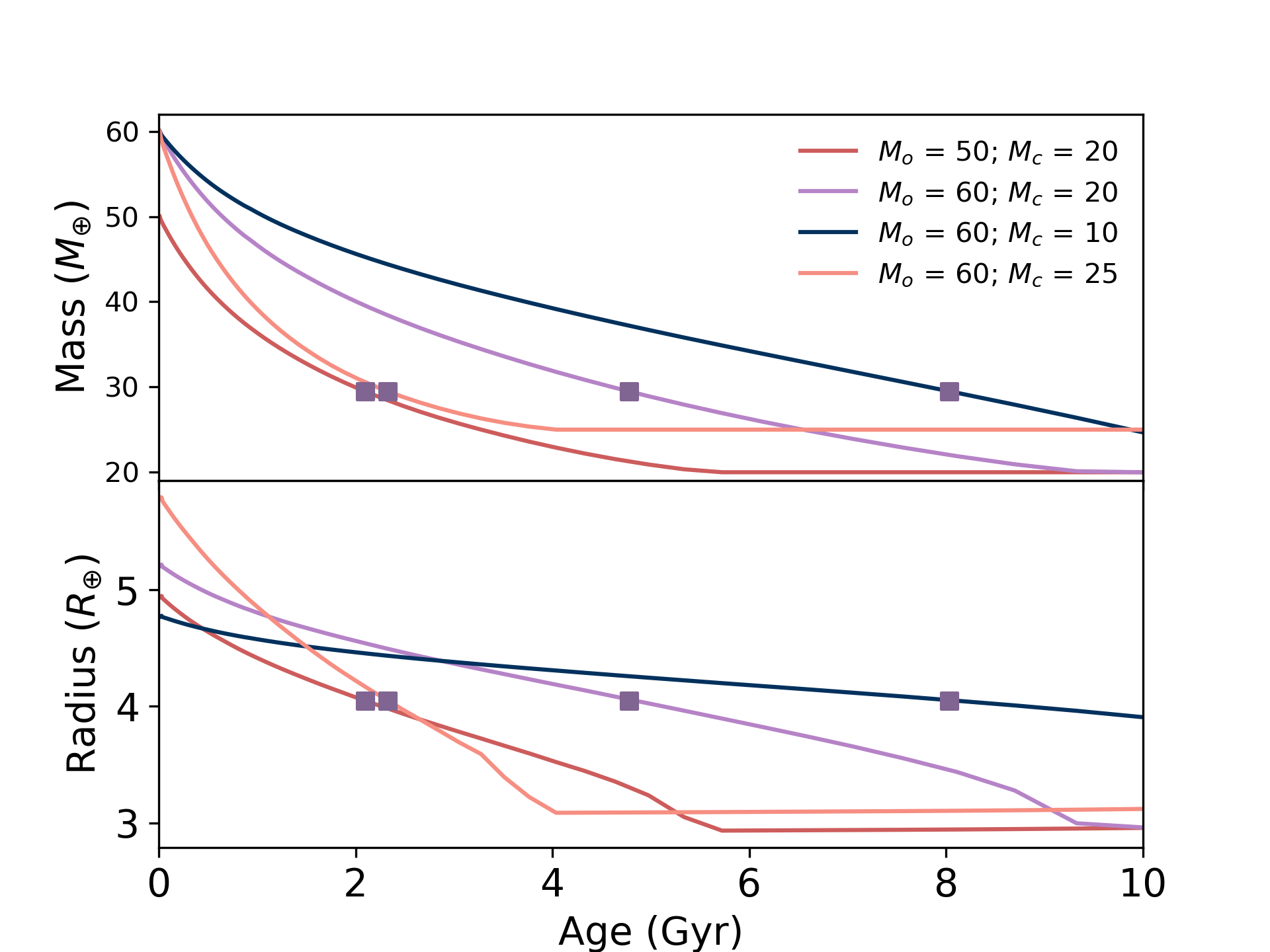}\hfill

\caption{Evolutionary tracks of TOI-3261b's mass (top panel) and radius (bottom panel), using models from \citet{Thorngren:2023}. These models assume an envelope enriched with metals. The planet's mass loss is modeled at various initial and core masses. The point at which the track passes through the observed mass and radius is indicated by the purple squares.}
\label{fig:massloss}

\end{figure}

 If we instead assume that metals are distributed from the core into the envelope, the envelope’s increased density then offers a greater resilience to escape, enabling TOI-3261b to retain its envelope across longer timescales with the same degree of XUV irradiation. Using planetary evolution simulations from \citet{Thorngren:2023}, we apply a similar approach to model TOI-3261b’s mass and radius evolution as a function of various core masses. However, unlike the previous simulations, in this case we do not assume an energy-limited mass loss rate, and we allow envelope metallicity to vary. Our initial conditions are generated by choosing a core mass and tuning the envelope metallicity so that the track intersects with the observed planet mass and radius. Generally, the envelope metallicity remains within the range of [Fe/H] = 0.6-0.8. The core is assumed to be isothermal, with no temperature discontinuity where the core meets the envelope and free transfer of energy across this boundary. These models, shown in Figure \ref{fig:massloss}, yield a lower limit of 50 {\mearth} for the planet's initial mass, with a core mass between 10-20 {\mearth}. 
 
 This model also finds that if we assume the system is {$\sim$}6.5 Gyr old, TOI-3261b's radius at 10 Myr old was no more than twice its current size. As higher initial masses require longer timescales to reach the current-day mass, photoevaporation could not reduce a Jupiter- or Saturn-sized planet into an object like TOI-3261b within the age of the universe. This means that if TOI-3261b did indeed originate as a gas giant, other mechanisms must be responsible for reducing its size.

\subsection{Tidal Stripping of a Progenitor Gas Giant}

To investigate the case where TOI-3261b formed as a giant planet, we must invoke more violent mass loss mechanisms. If the periastron distance was sufficiently reduced as a consequence of high eccentricity migration, a giant planet would fill its Roche lobe and undergo mass loss via Roche lobe overflow (RLO). As material is accreted by the star via RLO, the planet simultaneously shrinks in radius and migrates outward \citep{Valsecchi:2015}, halting the mass loss once it no longer fills its Roche lobe. In most cases, RLO completely strips the entire gaseous envelope, but more massive planets can retain a fraction of their envelopes to produce Neptune-sized objects like like TOI-3261b \citep{Liu:2013, Valsecchi:2015}. This is because as the planet contracts, its increased density provides increased resilience to mass loss and enables retention of its remaining envelope.

TOI-3261b’s observed location and suspected formerly-lower density are both conducive to triggering RLO. Following Eq. 1 of \citet{Liu:2013}, we find a present tidal disruption radius $r_t$ = 0.77 {\rsun} for TOI-3261b. The planet's current semimajor axis is $a_f$ $\sim$ 4.7 $r_t$ (or 3.5$r_{t,o}$, the initial tidal diruption radius). Assuming specific orbital angular momentum is conserved during the circularization process, its original periastron distance would have been approximately $r_{p,0}\sim a_f/2 = 1.75 r_{t,0}$. If we estimate that the progenitor of TOI-3261 b has a density similar to Jupiter ($\sim$1.0 $\mathrm{g}\,\mathrm{cm}^{-3}$), then the original tidal disruption radius is $r_{t,o}$ = 1.04 \rsun. Since $r_{p,o}$ / $r_{t,o}$ $<$ 2, if TOI-3261b was originally similar to Jupiter, it would have lost a large fraction of its mass upon its first periastron passage \citep{Faber:2005,Guillochon:2011}.

\begin{figure}[htp]
\centering
\includegraphics[width=.5\textwidth]{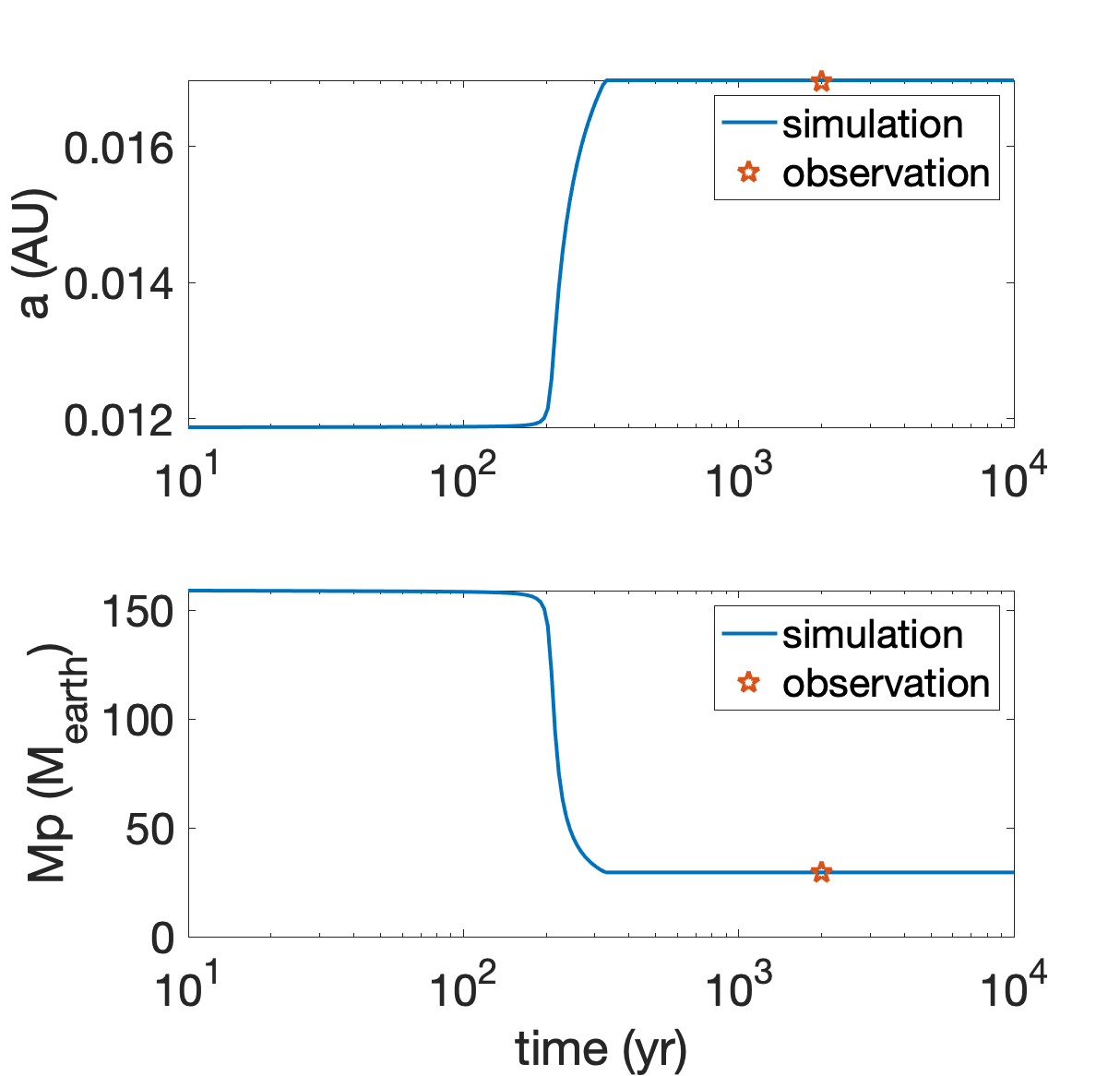}\hfill

\caption{Evolution of TOI-3261b's semimajor axis (top) and mass (bottom) as a result of tidal stripping via Roche Lobe Overflow, following the approach of \citet{Gu03}. The star denotes the planet's current position and mass.}
\label{fig:mesa}

\end{figure}

\begin{figure}[htp]
\centering
\includegraphics[width=.5\textwidth]{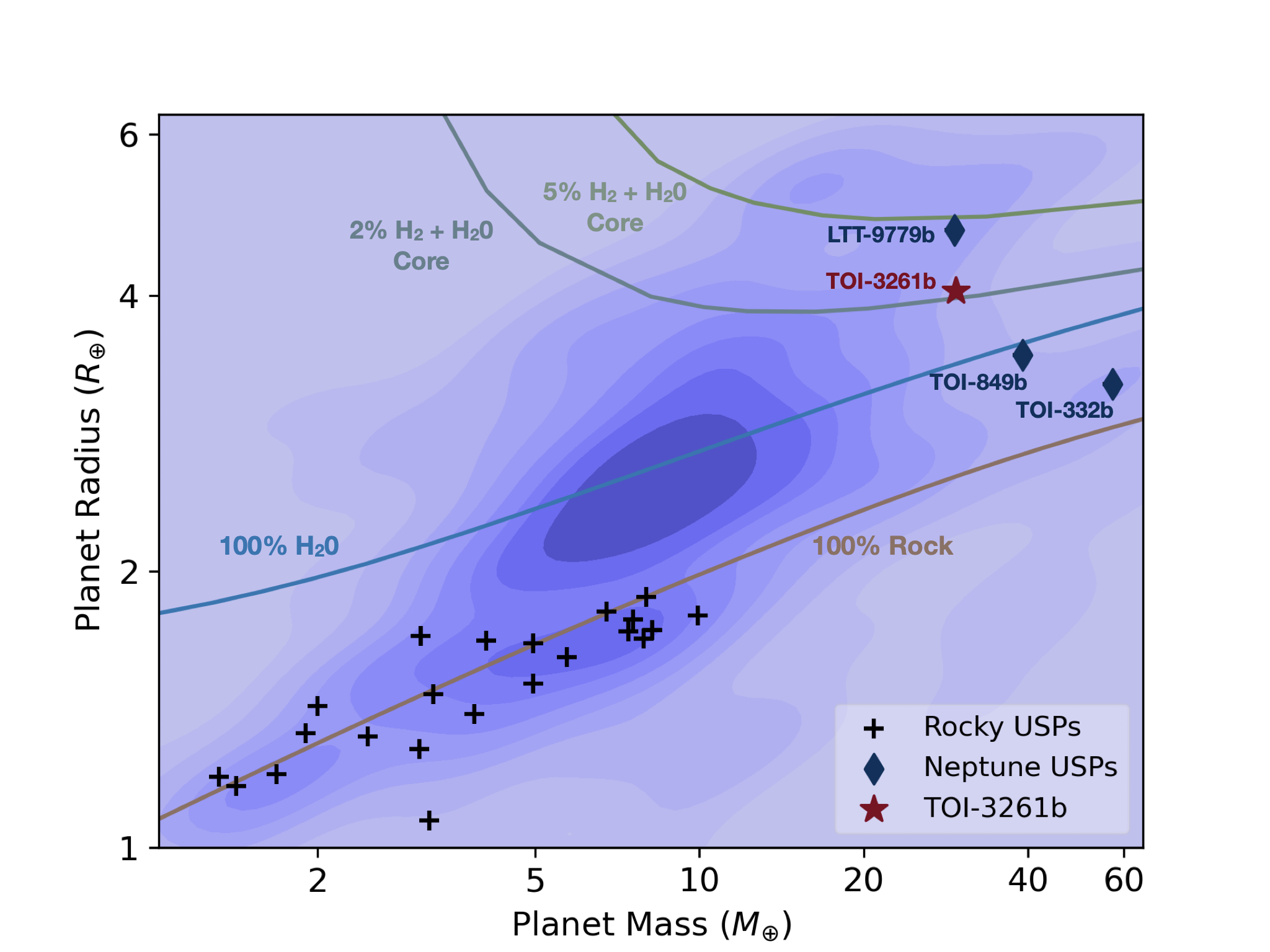}\hfill

\caption{Mass-radius distributions of ultrahot Neptunes compared to those of traditional USPs. The contour plot depicts the density in mass-radius space of the population of confirmed, published exoplanets listed in the Exoplanet Archive as of 2024 February 4. While the latter population is composed of pure rock, Neptune-sized USPs instead are expected to have a non-negligible envelope atop a water-rich core. Overplotted are mass-radius relations for pure rock, pure water, and a combined rock/water core with various {H$_2$} mass fractions. The mass-radius relations based on planet composition are taken from \citet{Zeng:2016}.}
\label{fig:massradius}

\end{figure}

We followed \citep{Gu03} to calculate the mass loss and orbital migration of TOI-3261b, assuming that the planet is a remnant core of a gas giant. Our simulation was initialized with a 0.5 Jupiter-mass planet with a core mass of 29.5 {\mearth}, at a distance of 0.012 AU. We track the changes in semimajor axis and planet mass due to RLO (Figure \ref{fig:mesa}), and find that this mechanism could produce an object at the same mass and location as TOI-3261b from a progenitor giant planet. This mechanism operates on shorter timescales than photoevaporation, and is potentially able to strip down a giant planet within a few Gyr \citep{Valsecchi:2014}.

The current mass and location of TOI-3261 b also match the final outcome from the adiabatic evolution of a Jupiter-mass planet experiencing mass loss with a 20 {\mearth} core \citep{Liu:2013}. The hydrodynamic simulations from \citet{Liu:2013} represent planets with cores as composite polytropes instead of the one-dimensional approximation implemented in the MESA models. This precisely models the dynamic response of planets during tidal interactions. Nevertheless, both of these approaches yield consistent results. Based on Figures 5 and 6 of \citet{Valsecchi:2015}, TOI-3261b's core mass coincides with a position between the 15 {\mearth} and 30 {\mearth} evolutionary curves, so that RLO culminates in the planet's current-day properties. 

\subsection{Future Observations}
Atmospheric analysis of TOI-3261b would reveal a rich vault of information, potentially applicable to not only the handful of Neptune-sized USPs, but extending towards the general population of hot giant planets. For instance, TOI-3261b's K-type host star is ideally suited for exciting the metastable 1083 nm helium line \citep{Spake:2018}. Investigating helium outflow from the planet could reveal the presence of a primordial atmosphere, as any existing helium would be escaping due to the planet's proximity to its star. In alignment with previous results for Neptune-sized planets, we expect helium outflow signals 5-10 times deeper than their transits in the optical regime. With a $\sim$1\% expected absorption, this signal would be detectable only with high-resolution infrared spectrographs in the Southern Hemisphere, like CRIRES+ on the European Southern Observatory's Very Large Telescope (VLT). Collecting multiple transits of TOI-3261b would allow unambiguous detection of helium outflow and distinguish between atmospheric composition scenarios. Conversely, a non-detection would therefore imply that TOI-3261b instead holds an enriched secondary atmosphere with a higher mean molecular weight, necessitating follow-up with more powerful instruments like the \textit{James Webb Space Telescope} (\textit{JWST}).

\textit{JWST} would offer an unprecedented view into the nature of TOI-3261 b. Using the framework established in \citet{Kempton:2018}, we derive an ESM of 12.02 for TOI-3261b, which matches that of TOI-332b and TOI-849b. This is higher than the recommended ESM value of 7.5 for benchmark TESS planets \citep{Kempton:2018}. LTT-9779b and TOI-849b have already been awarded time with \textit{JWST}/NIRSpec to obtain phase curves and eclipse measurements, respectively. The high ESM of TOI-3261b will allow us to determine if it has a high metallicity as predicted from a ``stripped core" scenario. A high metallicity scenario would then enable analysis of the planet's atmospheric dynamics with global circulation models, as was conducted with \textit{JWST} observations of GJ 1214b \citep{Kempton:2023, Gao:2023}. We would also be able to measure the C/O ratio of TOI-3261b, and infer the location of its formation.

TOI-3261b is one of the lowest-density USP Neptunes discovered to date, second only to LTT-9779b. This means that TOI-3261b provides a snapshot of an intermediate step in the envelope's mass loss, assuming that the end stage culminates in dense cores like TOI-849b and TOI-332b. The host star's negligible stellar activity also bolsters the viability of \textit{JWST} observations, as features like starspots can change the emission spectra of planets and increase uncertainties in their derived parameters \citep{Zellem:2017}.

Another avenue for follow-up would be measurement of TOI-3261b's obliquity through the Rossiter-McLaughlin (RM) effect. These observations would complement atmospheric data to create a more informed view of the system's history. If the planet is misaligned with its star, this gives evidence for violent dynamical processes earlier in its history such as impacts. However, the host star's faint magnitude, low $v\sin{i}$, and short transit duration present a significant challenge in measuring the stellar obliquity. From Equation 1 of \citet{Triaud2017}, we expect an RM semi-amplitude of {\bRM} m/s. Given that the precision we achieved with ESPRESSO with 15, 20-minute exposures was only $\sim$2 m/s, we would need to observe multiple transits of TOI-3261b to reach the necessary SNR. The resource-intensive programs required to confirm a 1 m/s signal may not be realistically feasible, as precision begins to be limited by the instrument's own intrinsic jitter.

No trends were detected across the {$\sim$}200-day baseline of the HARPS RV dataset, and periodograms show no signals suggesting additional planets with periods shorter than this baseline. However, companions with orbits on years-long timescales cannot be completely ruled out. For instance, assuming an orbital period of 1 year and a circular orbit with sin($i$) = 1, planets more massive than 0.53 {\mjup} would be detectable at the 3-{$\sigma$} level in the HARPS dataset. Any less massive companions would have semi-amplitudes too low to confirm, and the RV baseline is not long enough to reveal any trends that point to additional longer-period planets. Continued spectroscopic monitoring of TOI-3261 is imperative to confirm or deny the presence of any other planets, as their existence may reveal how TOI-3261b reached its current orbit. This would be a crucial way to either connect Neptune-sized USPs to a population of progenitor Hot Jupiters, or reveal an origin more similar to that of traditional USPs, who tend to reside in multiplanet systems \citep{Dai:2021}. So far, no companions to USP Neptunes have been found, suggesting that they are distinct from rocky USPs and potentially form through different means.

\section*{Acknowledgements}
E. N. acknowledges the PhD scholarship provided by the ARC discovery grant DP220100365.


This work makes use of observations from the LCOGT network. Part of the LCOGT telescope time was granted by NOIRLab through the Mid-Scale Innovations Program (MSIP). MSIP is funded by NSF.


This research has made use of the Exoplanet Follow-up Observation Program (\dataset[ExoFOP; DOI:10.26134/ExoFOP5]{https://doi.org/10.26134/ExoFOP5}) website, which is operated by the California Institute of Technology, under contract with the National Aeronautics and Space Administration under the Exoplanet Exploration Program.


Funding for the TESS mission is provided by NASA's Science Mission Directorate. KAC and CNW acknowledge support from the TESS mission via subaward s3449 from MIT.


Part of this research was carried out in part at the Jet Propulsion Laboratory, California Institute of Technology, under a contract with the National Aeronautics and Space Administration (80NM0018D0004).

This research was funded in part by the UKRI, (Grants ST/X001121/1, EP/X027562/1). Based on observations collected at the European Southern Observatory under ESO programme 108.21YY.001 (PI Armstrong).

The work of HPO has been carried out within the framework of the NCCR PlanetS supported by the Swiss National Science Foundation under grants 51NF40$\_$182901 and 51NF40$\_$205606.

X.D acknowledges the support from the European Research Council (ERC) under the European Union’s Horizon 2020 research and innovation programme (grant agreement SCORE No 851555) and from the Swiss National Science Foundation under the grant SPECTRE (No 200021$\_$215200)

J.L.-B. is funded by the MICIU/AEI/10.13039/501100011033 and NextGenerationEU/PRTR grants CNS2023-144309 and PID2019-107061GB-C61.

PJW acknowledges support from STFC under consolidated grants ST/T000406/1 and ST/X001121/1.

JSJ gratefully acknowledges support by FONDECYT grant 1240738 and from the ANID BASAL project FB210003.

This paper made use of data collected by the TESS mission and are publicly available from the Mikulski Archive for Space Telescopes (MAST) operated by the Space Telescope Science Institute (STScI). HLSP data from QLP can be accessed via \dataset[DOI:10.17909/t9-r086-e880]{https://doi.org/10.17909/t9-r086-e880}, and 2-minute SPOC data can be downloaded at \dataset[DOI:10.17909/t9-nmc8-f686]{https://doi.org/10.17909/t9-nmc8-f686}.
 
We acknowledge the use of public TESS data from pipelines at the TESS Science Office and at the TESS Science Processing Operations Center. 
 
Resources supporting this work were provided by the NASA High-End Computing (HEC) Program through the NASA Advanced Supercomputing (NAS) Division at Ames Research Center for the production of the SPOC data products.

\textit{Software:} {\tt AstroImageJ} \citep{Collins:2017}, {\tt batman} \citep{batman}, {\tt emcee} \citep{Foreman-Mackey2013}, {\tt isochrones} \citep{Isochrones}, {\tt matplotlib} \citep{Hunter2007}, {\tt pandas} \citep{McKinney2010}, {\tt RadVel} \citep{radvel}, {\tt scipy} \citep{Virtanen:2020}. 

\facilities{\textit{TESS}; LCOGT; ESO Paranal VLT 8.2m; ESO La Silla 3.6m; NASA Exoplanet Archive.}

\bibliographystyle{aasjournal}
\bibliography{refs}

\begin{thebibliography}{}
\expandafter\ifx\csname natexlab\endcsname\relax\def\natexlab#1{#1}\fi
\providecommand{\url}[1]{\href{#1}{#1}}
\providecommand{\dodoi}[1]{doi:~\href{http://doi.org/#1}{\nolinkurl{#1}}}
\providecommand{\doeprint}[1]{\href{http://ascl.net/#1}{\nolinkurl{http://ascl.net/#1}}}
\providecommand{\doarXiv}[1]{\href{https://arxiv.org/abs/#1}{\nolinkurl{https://arxiv.org/abs/#1}}}

\bibitem[{ESA(1997)}]{ESA1997}
 1997, ESA Special Publication, Vol. 1200, {The HIPPARCOS and TYCHO catalogues.
  Astrometric and photometric star catalogues derived from the ESA HIPPARCOS
  Space Astrometry Mission}

\bibitem[{{Adibekyan} {et~al.}(2015){Adibekyan}, {Figueira}, {Santos}, {Sousa},
  {Faria}, {Delgado-Mena}, {Oshagh}, {Tsantaki}, {Hakobyan}, {Gonz{\'a}lez
  Hern{\'a}ndez}, {Su{\'a}rez-Andr{\'e}s}, \& {Israelian}}]{Adibekyan-15}
{Adibekyan}, V., {Figueira}, P., {Santos}, N.~C., {et~al.} 2015, \aap, 583,
  A94, \dodoi{10.1051/0004-6361/201527120}

\bibitem[{{Adibekyan} {et~al.}(2012){Adibekyan}, {Sousa}, {Santos}, {Delgado
  Mena}, {Gonz{\'a}lez Hern{\'a}ndez}, {Israelian}, {Mayor}, \&
  {Khachatryan}}]{Adibekyan-12}
{Adibekyan}, V.~Z., {Sousa}, S.~G., {Santos}, N.~C., {et~al.} 2012, \aap, 545,
  A32, \dodoi{10.1051/0004-6361/201219401}

\bibitem[{{Allard} {et~al.}(2011){Allard}, {Homeier}, \&
  {Freytag}}]{Allard2011}
{Allard}, F., {Homeier}, D., \& {Freytag}, B. 2011, in Astronomical Society of
  the Pacific Conference Series, Vol. 448, 16th Cambridge Workshop on Cool
  Stars, Stellar Systems, and the Sun, ed. C.~{Johns-Krull}, M.~K. {Browning},
  \& A.~A. {West}, 91

\bibitem[{{Armstrong} {et~al.}(2020){Armstrong}, {Lopez}, {Adibekyan}, {Booth},
  {Bryant}, {Collins}, {Deleuil}, {Emsenhuber}, {Huang}, {King}, {Lillo-Box},
  {Lissauer}, {Matthews}, {Mousis}, {Nielsen}, {Osborn}, {Otegi}, {Santos},
  {Sousa}, {Stassun}, {Veras}, {Ziegler}, {Acton}, {Almenara}, {Anderson},
  {Barrado}, {Barros}, {Bayliss}, {Belardi}, {Bouchy}, {Brice{\~n}o}, {Brogi},
  {Brown}, {Burleigh}, {Casewell}, {Chaushev}, {Ciardi}, {Collins},
  {Col{\'o}n}, {Cooke}, {Crossfield}, {D{\'\i}az}, {Delgado Mena}, {Demangeon},
  {Dorn}, {Dumusque}, {Eigm{\"u}ller}, {Fausnaugh}, {Figueira}, {Gan},
  {Gandhi}, {Gill}, {Gonzales}, {Goad}, {G{\"u}nther}, {Helled}, {Hojjatpanah},
  {Howell}, {Jackman}, {Jenkins}, {Jenkins}, {Jensen}, {Kennedy}, {Latham},
  {Law}, {Lendl}, {Lozovsky}, {Mann}, {Moyano}, {McCormac}, {Meru},
  {Mordasini}, {Osborn}, {Pollacco}, {Queloz}, {Raynard}, {Ricker}, {Rowden},
  {Santerne}, {Schlieder}, {Seager}, {Sha}, {Tan}, {Tilbrook}, {Ting}, {Udry},
  {Vanderspek}, {Watson}, {West}, {Wilson}, {Winn}, {Wheatley}, {Villasenor},
  {Vines}, \& {Zhan}}]{Armstrong:2020}
{Armstrong}, D.~J., {Lopez}, T.~A., {Adibekyan}, V., {et~al.} 2020, \nat, 583,
  39, \dodoi{10.1038/s41586-020-2421-7}

\bibitem[{{Bailer-Jones} {et~al.}(2021){Bailer-Jones}, {Rybizki}, {Fouesneau},
  {Demleitner}, \& {Andrae}}]{BailerJones2021}
{Bailer-Jones}, C.~A.~L., {Rybizki}, J., {Fouesneau}, M., {Demleitner}, M., \&
  {Andrae}, R. 2021, \aj, 161, 147, \dodoi{10.3847/1538-3881/abd806}

\bibitem[{{Baranne} {et~al.}(1996){Baranne}, {Queloz}, {Mayor}, {Adrianzyk},
  {Knispel}, {Kohler}, {Lacroix}, {Meunier}, {Rimbaud}, \& {Vin}}]{Baranne1996}
{Baranne}, A., {Queloz}, D., {Mayor}, M., {et~al.} 1996, \aaps, 119, 373

\bibitem[{{Bianchi} {et~al.}(2017){Bianchi}, {Shiao}, \&
  {Thilker}}]{Bianchi2017}
{Bianchi}, L., {Shiao}, B., \& {Thilker}, D. 2017, \apjs, 230, 24,
  \dodoi{10.3847/1538-4365/aa7053}

\bibitem[{{Boisse} {et~al.}(2011){Boisse}, {Bouchy}, {H{\'e}brard}, {Bonfils},
  {Santos}, \& {Vauclair}}]{Boisse2011}
{Boisse}, I., {Bouchy}, F., {H{\'e}brard}, G., {et~al.} 2011, \aap, 528, A4,
  \dodoi{10.1051/0004-6361/201014354}

\bibitem[{{Brown} {et~al.}(2013){Brown}, {Baliber}, {Bianco}, {Bowman},
  {Burleson}, {Conway}, {Crellin}, {Depagne}, {De Vera}, {Dilday}, {Dragomir},
  {Dubberley}, {Eastman}, {Elphick}, {Falarski}, {Foale}, {Ford}, {Fulton},
  {Garza}, {Gomez}, {Graham}, {Greene}, {Haldeman}, {Hawkins}, {Haworth},
  {Haynes}, {Hidas}, {Hjelstrom}, {Howell}, {Hygelund}, {Lister}, {Lobdill},
  {Martinez}, {Mullins}, {Norbury}, {Parrent}, {Paulson}, {Petry}, {Pickles},
  {Posner}, {Rosing}, {Ross}, {Sand}, {Saunders}, {Shobbrook}, {Shporer},
  {Street}, {Thomas}, {Tsapras}, {Tufts}, {Valenti}, {Vander Horst}, {Walker},
  {White}, \& {Willis}}]{Brown:2013}
{Brown}, T.~M., {Baliber}, N., {Bianco}, F.~B., {et~al.} 2013, \pasp, 125,
  1031, \dodoi{10.1086/673168}

\bibitem[{{Castelli} \& {Kurucz}(2003)}]{Castelli2004}
{Castelli}, F., \& {Kurucz}, R.~L. 2003, in Modelling of Stellar Atmospheres,
  ed. N.~{Piskunov}, W.~W. {Weiss}, \& D.~F. {Gray}, Vol. 210, A20.
\newblock \doarXiv{astro-ph/0405087}

\bibitem[{{Chen} \& {Rogers}(2016)}]{ChenRogers2016}
{Chen}, H., \& {Rogers}, L.~A. 2016, \apj, 831, 180,
  \dodoi{10.3847/0004-637X/831/2/180}

\bibitem[{{Choi} {et~al.}(2016){Choi}, {Dotter}, {Conroy}, {Cantiello},
  {Paxton}, \& {Johnson}}]{MISTModels}
{Choi}, J., {Dotter}, A., {Conroy}, C., {et~al.} 2016, \apj, 823, 102,
  \dodoi{10.3847/0004-637X/823/2/102}

\bibitem[{{Collins}(2019)}]{collins:2019}
{Collins}, K. 2019, in American Astronomical Society Meeting Abstracts, Vol.
  233, American Astronomical Society Meeting Abstracts \#233, 140.05

\bibitem[{{Collins} {et~al.}(2017){Collins}, {Kielkopf}, {Stassun}, \&
  {Hessman}}]{Collins:2017}
{Collins}, K.~A., {Kielkopf}, J.~F., {Stassun}, K.~G., \& {Hessman}, F.~V.
  2017, \aj, 153, 77, \dodoi{10.3847/1538-3881/153/2/77}

\bibitem[{{Cutri} {et~al.}(2003){Cutri}, {Skrutskie}, {van Dyk}, {Beichman},
  {Carpenter}, {Chester}, {Cambresy}, {Evans}, {Fowler}, {Gizis}, {Howard},
  {Huchra}, {Jarrett}, {Kopan}, {Kirkpatrick}, {Light}, {Marsh}, {McCallon},
  {Schneider}, {Stiening}, {Sykes}, {Weinberg}, {Wheaton}, {Wheelock}, \&
  {Zacarias}}]{Cutri2003}
{Cutri}, R.~M., {Skrutskie}, M.~F., {van Dyk}, S., {et~al.} 2003, VizieR Online
  Data Catalog, II/246

\bibitem[{{Dai} {et~al.}(2021){Dai}, {Howard}, {Batalha}, {Beard}, {Behmard},
  {Blunt}, {Brinkman}, {Chontos}, {Crossfield}, {Dalba}, {Dressing}, {Fulton},
  {Giacalone}, {Hill}, {Huber}, {Isaacson}, {Kane}, {Lubin}, {Mayo},
  {Mo{\v{c}}nik}, {Akana Murphy}, {Petigura}, {Rice}, {Robertson}, {Rosenthal},
  {Roy}, {Rubenzahl}, {Weiss}, {Zandt}, {Beichman}, {Ciardi}, {Collins},
  {Gonzales}, {Howell}, {Matson}, {Matthews}, {Schlieder}, {Schwarz}, {Ricker},
  {Vanderspek}, {Latham}, {Seager}, {Winn}, {Jenkins}, {Caldwell}, {Colon},
  {Dragomir}, {Lund}, {McLean}, {Rudat}, \& {Shporer}}]{Dai:2021}
{Dai}, F., {Howard}, A.~W., {Batalha}, N.~M., {et~al.} 2021, \aj, 162, 62,
  \dodoi{10.3847/1538-3881/ac02bd}

\bibitem[{{Faber} {et~al.}(2005){Faber}, {Rasio}, \& {Willems}}]{Faber:2005}
{Faber}, J.~A., {Rasio}, F.~A., \& {Willems}, B. 2005, \icarus, 175, 248,
  \dodoi{10.1016/j.icarus.2004.10.021}

\bibitem[{{Fern{\'a}ndez Fern{\'a}ndez} {et~al.}(2023){Fern{\'a}ndez
  Fern{\'a}ndez}, {Wheatley}, \& {King}}]{Fernandez-Fernandez:2023}
{Fern{\'a}ndez Fern{\'a}ndez}, J., {Wheatley}, P.~J., \& {King}, G.~W. 2023,
  \mnras, 522, 4251, \dodoi{10.1093/mnras/stad1257}

\bibitem[{{Fischer} \& {Valenti}(2005)}]{FischerValenti2005}
{Fischer}, D.~A., \& {Valenti}, J. 2005, \apj, 622, 1102,
  \dodoi{10.1086/428383}

\bibitem[{{Foreman-Mackey} {et~al.}(2013{\natexlab{a}}){Foreman-Mackey},
  {Hogg}, {Lang}, \& {Goodman}}]{ForemanMackey:2012}
{Foreman-Mackey}, D., {Hogg}, D.~W., {Lang}, D., \& {Goodman}, J.
  2013{\natexlab{a}}, \pasp, 125, 306, \dodoi{10.1086/670067}

\bibitem[{{Foreman-Mackey} {et~al.}(2013{\natexlab{b}}){Foreman-Mackey},
  {Hogg}, {Lang}, \& {Goodman}}]{Foreman-Mackey2013}
---. 2013{\natexlab{b}}, \pasp, 125, 306, \dodoi{10.1086/670067}

\bibitem[{{Fulton} {et~al.}(2018{\natexlab{a}}){Fulton}, {Petigura}, {Blunt},
  \& {Sinukoff}}]{Fulton2018}
{Fulton}, B.~J., {Petigura}, E.~A., {Blunt}, S., \& {Sinukoff}, E.
  2018{\natexlab{a}}, \pasp, 130, 044504, \dodoi{10.1088/1538-3873/aaaaa8}

\bibitem[{{Fulton} {et~al.}(2018{\natexlab{b}}){Fulton}, {Petigura}, {Blunt},
  \& {Sinukoff}}]{radvel}
---. 2018{\natexlab{b}}, \pasp, 130, 044504, \dodoi{10.1088/1538-3873/aaaaa8}

\bibitem[{{Gaia Collaboration} {et~al.}(2016){Gaia Collaboration}, {Prusti},
  {de Bruijne}, {Brown}, {Vallenari}, {Babusiaux}, {Bailer-Jones}, {Bastian},
  {Biermann}, {Evans}, {Eyer}, {Jansen}, {Jordi}, {Klioner}, {Lammers},
  {Lindegren}, {Luri}, {Mignard}, {Milligan}, {Panem}, {Poinsignon},
  {Pourbaix}, {Randich}, {Sarri}, {Sartoretti}, {Siddiqui}, {Soubiran},
  {Valette}, {van Leeuwen}, {Walton}, {Aerts}, {Arenou}, {Cropper}, {Drimmel},
  {H{\o}g}, {Katz}, {Lattanzi}, {O'Mullane}, {Grebel}, {Holland}, {Huc},
  {Passot}, {Bramante}, {Cacciari}, {Casta{\~n}eda}, {Chaoul}, {Cheek}, {De
  Angeli}, {Fabricius}, {Guerra}, {Hern{\'a}ndez}, {Jean-Antoine-Piccolo},
  {Masana}, {Messineo}, {Mowlavi}, {Nienartowicz}, {Ord{\'o}{\~n}ez-Blanco},
  {Panuzzo}, {Portell}, {Richards}, {Riello}, {Seabroke}, {Tanga},
  {Th{\'e}venin}, {Torra}, {Els}, {Gracia-Abril}, {Comoretto},
  {Garcia-Reinaldos}, {Lock}, {Mercier}, {Altmann}, {Andrae}, {Astraatmadja},
  {Bellas-Velidis}, {Benson}, {Berthier}, {Blomme}, {Busso}, {Carry},
  {Cellino}, {Clementini}, {Cowell}, {Creevey}, {Cuypers}, {Davidson}, {De
  Ridder}, {de Torres}, {Delchambre}, {Dell'Oro}, {Ducourant}, {Fr{\'e}mat},
  {Garc{\'\i}a-Torres}, {Gosset}, {Halbwachs}, {Hambly}, {Harrison}, {Hauser},
  {Hestroffer}, {Hodgkin}, {Huckle}, {Hutton}, {Jasniewicz}, {Jordan},
  {Kontizas}, {Korn}, {Lanzafame}, {Manteiga}, {Moitinho}, {Muinonen},
  {Osinde}, {Pancino}, {Pauwels}, {Petit}, {Recio-Blanco}, {Robin}, {Sarro},
  {Siopis}, {Smith}, {Smith}, {Sozzetti}, {Thuillot}, {van Reeven}, {Viala},
  {Abbas}, {Abreu Aramburu}, {Accart}, {Aguado}, {Allan}, {Allasia},
  {Altavilla}, {{\'A}lvarez}, {Alves}, {Anderson}, {Andrei}, {Anglada Varela},
  {Antiche}, {Antoja}, {Ant{\'o}n}, {Arcay}, {Atzei}, {Ayache}, {Bach},
  {Baker}, {Balaguer-N{\'u}{\~n}ez}, {Barache}, {Barata}, {Barbier}, {Barblan},
  {Baroni}, {Barrado y Navascu{\'e}s}, {Barros}, {Barstow}, {Becciani},
  {Bellazzini}, {Bellei}, {Bello Garc{\'\i}a}, {Belokurov}, {Bendjoya},
  {Berihuete}, {Bianchi}, {Bienaym{\'e}}, {Billebaud}, {Blagorodnova},
  {Blanco-Cuaresma}, {Boch}, {Bombrun}, {Borrachero}, {Bouquillon}, {Bourda},
  {Bouy}, {Bragaglia}, {Breddels}, {Brouillet}, {Br{\"u}semeister},
  {Bucciarelli}, {Budnik}, {Burgess}, {Burgon}, {Burlacu}, {Busonero}, {Buzzi},
  {Caffau}, {Cambras}, {Campbell}, {Cancelliere}, {Cantat-Gaudin}, {Carlucci},
  {Carrasco}, {Castellani}, {Charlot}, {Charnas}, {Charvet}, {Chassat},
  {Chiavassa}, {Clotet}, {Cocozza}, {Collins}, {Collins}, {Costigan}, {Crifo},
  {Cross}, {Crosta}, {Crowley}, {Dafonte}, {Damerdji}, {Dapergolas}, {David},
  {David}, {De Cat}, {de Felice}, {de Laverny}, {De Luise}, {De March}, {de
  Martino}, {de Souza}, {Debosscher}, {del Pozo}, {Delbo}, {Delgado},
  {Delgado}, {di Marco}, {Di Matteo}, {Diakite}, {Distefano}, {Dolding}, {Dos
  Anjos}, {Drazinos}, {Dur{\'a}n}, {Dzigan}, {Ecale}, {Edvardsson}, {Enke},
  {Erdmann}, {Escolar}, {Espina}, {Evans}, {Eynard Bontemps}, {Fabre},
  {Fabrizio}, {Faigler}, {Falc{\~a}o}, {Farr{\`a}s Casas}, {Faye}, {Federici},
  {Fedorets}, {Fern{\'a}ndez-Hern{\'a}ndez}, {Fernique}, {Fienga}, {Figueras},
  {Filippi}, {Findeisen}, {Fonti}, {Fouesneau}, {Fraile}, {Fraser}, {Fuchs},
  {Furnell}, {Gai}, {Galleti}, {Galluccio}, {Garabato}, {Garc{\'\i}a-Sedano},
  {Gar{\'e}}, {Garofalo}, {Garralda}, {Gavras}, {Gerssen}, {Geyer}, {Gilmore},
  {Girona}, {Giuffrida}, {Gomes}, {Gonz{\'a}lez-Marcos},
  {Gonz{\'a}lez-N{\'u}{\~n}ez}, {Gonz{\'a}lez-Vidal}, {Granvik}, {Guerrier},
  {Guillout}, {Guiraud}, {G{\'u}rpide}, {Guti{\'e}rrez-S{\'a}nchez}, {Guy},
  {Haigron}, {Hatzidimitriou}, {Haywood}, {Heiter}, {Helmi}, {Hobbs},
  {Hofmann}, {Holl}, {Holland}, {Hunt}, {Hypki}, {Icardi}, {Irwin}, {Jevardat
  de Fombelle}, {Jofr{\'e}}, {Jonker}, {Jorissen}, {Julbe}, {Karampelas},
  {Kochoska}, {Kohley}, {Kolenberg}, {Kontizas}, {Koposov}, {Kordopatis},
  {Koubsky}, {Kowalczyk}, {Krone-Martins}, {Kudryashova}, {Kull}, {Bachchan},
  {Lacoste-Seris}, {Lanza}, {Lavigne}, {Le Poncin-Lafitte}, {Lebreton},
  {Lebzelter}, {Leccia}, {Leclerc}, {Lecoeur-Taibi}, {Lemaitre}, {Lenhardt},
  {Leroux}, {Liao}, {Licata}, {Lindstr{\o}m}, {Lister}, {Livanou}, {Lobel},
  {L{\"o}ffler}, {L{\'o}pez}, {Lopez-Lozano}, {Lorenz}, {Loureiro},
  {MacDonald}, {Magalh{\~a}es Fernandes}, {Managau}, {Mann}, {Mantelet},
  {Marchal}, {Marchant}, {Marconi}, {Marie}, {Marinoni}, {Marrese},
  {Marschalk{\'o}}, {Marshall}, {Mart{\'\i}n-Fleitas}, {Martino}, {Mary},
  {Matijevi{\v{c}}}, {Mazeh}, {McMillan}, {Messina}, {Mestre}, {Michalik},
  {Millar}, {Miranda}, {Molina}, {Molinaro}, {Molinaro}, {Moln{\'a}r},
  {Moniez}, {Montegriffo}, {Monteiro}, {Mor}, {Mora}, {Morbidelli}, {Morel},
  {Morgenthaler}, {Morley}, {Morris}, {Mulone}, {Muraveva}, {Musella},
  {Narbonne}, {Nelemans}, {Nicastro}, {Noval}, {Ord{\'e}novic},
  {Ordieres-Mer{\'e}}, {Osborne}, {Pagani}, {Pagano}, {Pailler}, {Palacin},
  {Palaversa}, {Parsons}, {Paulsen}, {Pecoraro}, {Pedrosa}, {Pentik{\"a}inen},
  {Pereira}, {Pichon}, {Piersimoni}, {Pineau}, {Plachy}, {Plum}, {Poujoulet},
  {Pr{\v{s}}a}, {Pulone}, {Ragaini}, {Rago}, {Rambaux}, {Ramos-Lerate},
  {Ranalli}, {Rauw}, {Read}, {Regibo}, {Renk}, {Reyl{\'e}}, {Ribeiro},
  {Rimoldini}, {Ripepi}, {Riva}, {Rixon}, {Roelens}, {Romero-G{\'o}mez},
  {Rowell}, {Royer}, {Rudolph}, {Ruiz-Dern}, {Sadowski}, {Sagrist{\`a}
  Sell{\'e}s}, {Sahlmann}, {Salgado}, {Salguero}, {Sarasso}, {Savietto},
  {Schnorhk}, {Schultheis}, {Sciacca}, {Segol}, {Segovia}, {Segransan},
  {Serpell}, {Shih}, {Smareglia}, {Smart}, {Smith}, {Solano}, {Solitro},
  {Sordo}, {Soria Nieto}, {Souchay}, {Spagna}, {Spoto}, {Stampa}, {Steele},
  {Steidelm{\"u}ller}, {Stephenson}, {Stoev}, {Suess}, {S{\"u}veges}, {Surdej},
  {Szabados}, {Szegedi-Elek}, {Tapiador}, {Taris}, {Tauran}, {Taylor},
  {Teixeira}, {Terrett}, {Tingley}, {Trager}, {Turon}, {Ulla}, {Utrilla},
  {Valentini}, {van Elteren}, {Van Hemelryck}, {van Leeuwen}, {Varadi},
  {Vecchiato}, {Veljanoski}, {Via}, {Vicente}, {Vogt}, {Voss}, {Votruba},
  {Voutsinas}, {Walmsley}, {Weiler}, {Weingrill}, {Werner}, {Wevers},
  {Whitehead}, {Wyrzykowski}, {Yoldas}, {{\v{Z}}erjal}, {Zucker}, {Zurbach},
  {Zwitter}, {Alecu}, {Allen}, {Allende Prieto}, {Amorim},
  {Anglada-Escud{\'e}}, {Arsenijevic}, {Azaz}, {Balm}, {Beck}, {Bernstein},
  {Bigot}, {Bijaoui}, {Blasco}, {Bonfigli}, {Bono}, {Boudreault}, {Bressan},
  {Brown}, {Brunet}, {Bunclark}, {Buonanno}, {Butkevich}, {Carret}, {Carrion},
  {Chemin}, {Ch{\'e}reau}, {Corcione}, {Darmigny}, {de Boer}, {de Teodoro}, {de
  Zeeuw}, {Delle Luche}, {Domingues}, {Dubath}, {Fodor}, {Fr{\'e}zouls},
  {Fries}, {Fustes}, {Fyfe}, {Gallardo}, {Gallegos}, {Gardiol}, {Gebran},
  {Gomboc}, {G{\'o}mez}, {Grux}, {Gueguen}, {Heyrovsky}, {Hoar}, {Iannicola},
  {Isasi Parache}, {Janotto}, {Joliet}, {Jonckheere}, {Keil}, {Kim},
  {Klagyivik}, {Klar}, {Knude}, {Kochukhov}, {Kolka}, {Kos}, {Kutka}, {Lainey},
  {LeBouquin}, {Liu}, {Loreggia}, {Makarov}, {Marseille}, {Martayan},
  {Martinez-Rubi}, {Massart}, {Meynadier}, {Mignot}, {Munari}, {Nguyen},
  {Nordlander}, {Ocvirk}, {O'Flaherty}, {Olias Sanz}, {Ortiz}, {Osorio},
  {Oszkiewicz}, {Ouzounis}, {Palmer}, {Park}, {Pasquato}, {Peltzer}, {Peralta},
  {P{\'e}turaud}, {Pieniluoma}, {Pigozzi}, {Poels}, {Prat}, {Prod'homme},
  {Raison}, {Rebordao}, {Risquez}, {Rocca-Volmerange}, {Rosen}, {Ruiz-Fuertes},
  {Russo}, {Sembay}, {Serraller Vizcaino}, {Short}, {Siebert}, {Silva},
  {Sinachopoulos}, {Slezak}, {Soffel}, {Sosnowska}, {Strai{\v{z}}ys}, {ter
  Linden}, {Terrell}, {Theil}, {Tiede}, {Troisi}, {Tsalmantza}, {Tur},
  {Vaccari}, {Vachier}, {Valles}, {Van Hamme}, {Veltz}, {Virtanen}, {Wallut},
  {Wichmann}, {Wilkinson}, {Ziaeepour}, \& {Zschocke}}]{Gaia}
{Gaia Collaboration}, {Prusti}, T., {de Bruijne}, J.~H.~J., {et~al.} 2016,
  \aap, 595, A1, \dodoi{10.1051/0004-6361/201629272}

\bibitem[{{Gaia Collaboration} {et~al.}(2022){Gaia Collaboration}, {Vallenari},
  {Brown}, {Prusti}, {de Bruijne}, {Arenou}, {Babusiaux}, {Biermann},
  {Creevey}, {Ducourant}, {Evans}, {Eyer}, {Guerra}, {Hutton}, {Jordi},
  {Klioner}, {Lammers}, {Lindegren}, {Luri}, {Mignard}, {Panem}, {Pourbaix},
  {Randich}, {Sartoretti}, {Soubiran}, {Tanga}, {Walton}, {Bailer-Jones},
  {Bastian}, {Drimmel}, {Jansen}, {Katz}, {Lattanzi}, {van Leeuwen}, {Bakker},
  {Cacciari}, {Casta{\~n}eda}, {De Angeli}, {Fabricius}, {Fouesneau},
  {Fr{\'e}mat}, {Galluccio}, {Guerrier}, {Heiter}, {Masana}, {Messineo},
  {Mowlavi}, {Nicolas}, {Nienartowicz}, {Pailler}, {Panuzzo}, {Riclet}, {Roux},
  {Seabroke}, {Sordo{\o}rcit}, {Th{\'e}venin}, {Gracia-Abril}, {Portell},
  {Teyssier}, {Altmann}, {Andrae}, {Audard}, {Bellas-Velidis}, {Benson},
  {Berthier}, {Blomme}, {Burgess}, {Busonero}, {Busso}, {C{\'a}novas}, {Carry},
  {Cellino}, {Cheek}, {Clementini}, {Damerdji}, {Davidson}, {de Teodoro},
  {Nu{\~n}ez Campos}, {Delchambre}, {Dell'Oro}, {Esquej},
  {Fern{\'a}ndez-Hern{\'a}ndez}, {Fraile}, {Garabato}, {Garc{\'\i}a-Lario},
  {Gosset}, {Haigron}, {Halbwachs}, {Hambly}, {Harrison}, {Hern{\'a}ndez},
  {Hestroffer}, {Hodgkin}, {Holl}, {Jan{\ss}en}, {Jevardat de Fombelle},
  {Jordan}, {Krone-Martins}, {Lanzafame}, {L{\"o}ffler}, {Marchal}, {Marrese},
  {Moitinho}, {Muinonen}, {Osborne}, {Pancino}, {Pauwels}, {Recio-Blanco},
  {Reyl{\'e}}, {Riello}, {Rimoldini}, {Roegiers}, {Rybizki}, {Sarro}, {Siopis},
  {Smith}, {Sozzetti}, {Utrilla}, {van Leeuwen}, {Abbas}, {{\'A}brah{\'a}m},
  {Abreu Aramburu}, {Aerts}, {Aguado}, {Ajaj}, {Aldea-Montero}, {Altavilla},
  {{\'A}lvarez}, {Alves}, {Anders}, {Anderson}, {Anglada Varela}, {Antoja},
  {Baines}, {Baker}, {Balaguer-N{\'u}{\~n}ez}, {Balbinot}, {Balog}, {Barache},
  {Barbato}, {Barros}, {Barstow}, {Bartolom{\'e}}, {Bassilana}, {Bauchet},
  {Becciani}, {Bellazzini}, {Berihuete}, {Bernet}, {Bertone}, {Bianchi},
  {Binnenfeld}, {Blanco-Cuaresma}, {Blazere}, {Boch}, {Bombrun}, {Bossini},
  {Bouquillon}, {Bragaglia}, {Bramante}, {Breedt}, {Bressan}, {Brouillet},
  {Brugaletta}, {Bucciarelli}, {Burlacu}, {Butkevich}, {Buzzi}, {Caffau},
  {Cancelliere}, {Cantat-Gaudin}, {Carballo}, {Carlucci}, {Carnerero},
  {Carrasco}, {Casamiquela}, {Castellani}, {Castro-Ginard}, {Chaoul},
  {Charlot}, {Chemin}, {Chiaramida}, {Chiavassa}, {Chornay}, {Comoretto},
  {Contursi}, {Cooper}, {Cornez}, {Cowell}, {Crifo}, {Cropper}, {Crosta},
  {Crowley}, {Dafonte}, {Dapergolas}, {David}, {David}, {de Laverny}, {De
  Luise}, {De March}, {De Ridder}, {de Souza}, {de Torres}, {del Peloso}, {del
  Pozo}, {Delbo}, {Delgado}, {Delisle}, {Demouchy}, {Dharmawardena}, {Di
  Matteo}, {Diakite}, {Diener}, {Distefano}, {Dolding}, {Edvardsson}, {Enke},
  {Fabre}, {Fabrizio}, {Faigler}, {Fedorets}, {Fernique}, {Fienga}, {Figueras},
  {Fournier}, {Fouron}, {Fragkoudi}, {Gai}, {Garcia-Gutierrez},
  {Garcia-Reinaldos}, {Garc{\'\i}a-Torres}, {Garofalo}, {Gavel}, {Gavras},
  {Gerlach}, {Geyer}, {Giacobbe}, {Gilmore}, {Girona}, {Giuffrida}, {Gomel},
  {Gomez}, {Gonz{\'a}lez-N{\'u}{\~n}ez}, {Gonz{\'a}lez-Santamar{\'\i}a},
  {Gonz{\'a}lez-Vidal}, {Granvik}, {Guillout}, {Guiraud},
  {Guti{\'e}rrez-S{\'a}nchez}, {Guy}, {Hatzidimitriou}, {Hauser}, {Haywood},
  {Helmer}, {Helmi}, {Sarmiento}, {Hidalgo}, {Hilger}, {H{\l}adczuk}, {Hobbs},
  {Holland}, {Huckle}, {Jardine}, {Jasniewicz}, {Jean-Antoine Piccolo},
  {Jim{\'e}nez-Arranz}, {Jorissen}, {Juaristi Campillo}, {Julbe}, {Karbevska},
  {Kervella}, {Khanna}, {Kontizas}, {Kordopatis}, {Korn}, {K{\'o}sp{\'a}l},
  {Kostrzewa-Rutkowska}, {Kruszy{\'n}ska}, {Kun}, {Laizeau}, {Lambert},
  {Lanza}, {Lasne}, {Le Campion}, {Lebreton}, {Lebzelter}, {Leccia}, {Leclerc},
  {Lecoeur-Taibi}, {Liao}, {Licata}, {Lindstr{\o}m}, {Lister}, {Livanou},
  {Lobel}, {Lorca}, {Loup}, {Madrero Pardo}, {Magdaleno Romeo}, {Managau},
  {Mann}, {Manteiga}, {Marchant}, {Marconi}, {Marcos}, {Marcos Santos},
  {Mar{\'\i}n Pina}, {Marinoni}, {Marocco}, {Marshall}, {Polo},
  {Mart{\'\i}n-Fleitas}, {Marton}, {Mary}, {Masip}, {Massari},
  {Mastrobuono-Battisti}, {Mazeh}, {McMillan}, {Messina}, {Michalik}, {Millar},
  {Mints}, {Molina}, {Molinaro}, {Moln{\'a}r}, {Monari}, {Mongui{\'o}},
  {Montegriffo}, {Montero}, {Mor}, {Mora}, {Morbidelli}, {Morel}, {Morris},
  {Muraveva}, {Murphy}, {Musella}, {Nagy}, {Noval}, {Oca{\~n}a}, {Ogden},
  {Ordenovic}, {Osinde}, {Pagani}, {Pagano}, {Palaversa}, {Palicio},
  {Pallas-Quintela}, {Panahi}, {Payne-Wardenaar}, {Pe{\~n}alosa Esteller},
  {Penttil{\"a}}, {Pichon}, {Piersimoni}, {Pineau}, {Plachy}, {Plum}, {Poggio},
  {Pr{\v{s}}a}, {Pulone}, {Racero}, {Ragaini}, {Rainer}, {Raiteri}, {Rambaux},
  {Ramos}, {Ramos-Lerate}, {Re Fiorentin}, {Regibo}, {Richards}, {Rios Diaz},
  {Ripepi}, {Riva}, {Rix}, {Rixon}, {Robichon}, {Robin}, {Robin}, {Roelens},
  {Rogues}, {Rohrbasser}, {Romero-G{\'o}mez}, {Rowell}, {Royer}, {Ruz Mieres},
  {Rybicki}, {Sadowski}, {S{\'a}ez N{\'u}{\~n}ez}, {Sagrist{\`a} Sell{\'e}s},
  {Sahlmann}, {Salguero}, {Samaras}, {Sanchez Gimenez}, {Sanna},
  {Santove{\~n}a}, {Sarasso}, {Schultheis}, {Sciacca}, {Segol}, {Segovia},
  {S{\'e}gransan}, {Semeux}, {Shahaf}, {Siddiqui}, {Siebert}, {Siltala},
  {Silvelo}, {Slezak}, {Slezak}, {Smart}, {Snaith}, {Solano}, {Solitro},
  {Souami}, {Souchay}, {Spagna}, {Spina}, {Spoto}, {Steele},
  {Steidelm{\"u}ller}, {Stephenson}, {S{\"u}veges}, {Surdej}, {Szabados},
  {Szegedi-Elek}, {Taris}, {Taylo}, {Teixeira}, {Tolomei}, {Tonello}, {Torra},
  {Torra}, {Torralba Elipe}, {Trabucchi}, {Tsounis}, {Turon}, {Ulla}, {Unger},
  {Vaillant}, {van Dillen}, {van Reeven}, {Vanel}, {Vecchiato}, {Viala},
  {Vicente}, {Voutsinas}, {Weiler}, {Wevers}, {Wyrzykowski}, {Yoldas}, {Yvard},
  {Zhao}, {Zorec}, {Zucker}, \& {Zwitter}}]{GaiaDR3}
{Gaia Collaboration}, {Vallenari}, A., {Brown}, A.~G.~A., {et~al.} 2022, arXiv
  e-prints, arXiv:2208.00211.
\newblock \doarXiv{2208.00211}

\bibitem[{{Gao} {et~al.}(2023){Gao}, {Piette}, {Steinrueck}, {Nixon}, {Zhang},
  {Kempton}, {Bean}, {Rauscher}, {Parmentier}, {Batalha}, {Savel}, {Arnold},
  {Roman}, {Malsky}, \& {Taylor}}]{Gao:2023}
{Gao}, P., {Piette}, A. A.~A., {Steinrueck}, M.~E., {et~al.} 2023, \apj, 951,
  96, \dodoi{10.3847/1538-4357/acd16f}

\bibitem[{{Gomes da Silva} {et~al.}(2018){Gomes da Silva}, {Figueira},
  {Santos}, \& {Faria}}]{GomesdaSilva:2018}
{Gomes da Silva}, J., {Figueira}, P., {Santos}, N., \& {Faria}, J. 2018, The
  Journal of Open Source Software, 3, 667, \dodoi{10.21105/joss.00667}

\bibitem[{{Gomes da Silva} {et~al.}(2021){Gomes da Silva}, {Santos},
  {Adibekyan}, {Sousa}, {Campante}, {Figueira}, {Bossini}, {Delgado-Mena},
  {Monteiro}, {de Laverny}, {Recio-Blanco}, \& {Lovis}}]{GomesdaSilva:2021}
{Gomes da Silva}, J., {Santos}, N.~C., {Adibekyan}, V., {et~al.} 2021, \aap,
  646, A77, \dodoi{10.1051/0004-6361/202039765}

\bibitem[{{Gu} {et~al.}(2003){Gu}, {Lin}, \& {Bodenheimer}}]{Gu03}
{Gu}, P.-G., {Lin}, D. N.~C., \& {Bodenheimer}, P.~H. 2003, \apj, 588, 509,
  \dodoi{10.1086/373920}

\bibitem[{{Guillochon} {et~al.}(2011){Guillochon}, {Ramirez-Ruiz}, \&
  {Lin}}]{Guillochon:2011}
{Guillochon}, J., {Ramirez-Ruiz}, E., \& {Lin}, D. 2011, \apj, 732, 74,
  \dodoi{10.1088/0004-637X/732/2/74}

\bibitem[{{Huang} {et~al.}(2020{\natexlab{a}}){Huang}, {Vanderburg}, {P{\'a}l},
  {Sha}, {Yu}, {Fong}, {Fausnaugh}, {Shporer}, {Guerrero}, {Vanderspek}, \&
  {Ricker}}]{Huang:2020b}
{Huang}, C.~X., {Vanderburg}, A., {P{\'a}l}, A., {et~al.} 2020{\natexlab{a}},
  Research Notes of the American Astronomical Society, 4, 206,
  \dodoi{10.3847/2515-5172/abca2d}

\bibitem[{{Huang} {et~al.}(2020{\natexlab{b}}){Huang}, {Vanderburg}, {P{\'a}l},
  {Sha}, {Yu}, {Fong}, {Fausnaugh}, {Shporer}, {Guerrero}, {Vanderspek}, \&
  {Ricker}}]{Huang:2020a}
---. 2020{\natexlab{b}}, Research Notes of the American Astronomical Society,
  4, 204, \dodoi{10.3847/2515-5172/abca2e}

\bibitem[{Hunter(2007)}]{Hunter2007}
Hunter, J.~D. 2007, Computing in Science \& Engineering, 9, 90,
  \dodoi{10.1109/MCSE.2007.55}

\bibitem[{Husser {et~al.}(2013)Husser, {Wende-von Berg}, Dreizler, Homeier,
  Reiners, Barman, \& Hauschildt}]{Husser2013}
Husser, T.-O., {Wende-von Berg}, S., Dreizler, S., {et~al.} 2013, A{\&}A, 553,
  A6, \dodoi{10.1051/0004-6361/201219058}

\bibitem[{{Jenkins}(2002)}]{Jenkins:2002}
{Jenkins}, J.~M. 2002, \apj, 575, 493, \dodoi{10.1086/341136}

\bibitem[{{Jenkins} {et~al.}(2020{\natexlab{a}}){Jenkins}, {Tenenbaum},
  {Seader}, {Burke}, {McCauliff}, {Smith}, {Twicken}, \&
  {Chandrasekaran}}]{SPOC2020}
{Jenkins}, J.~M., {Tenenbaum}, P., {Seader}, S., {et~al.} 2020{\natexlab{a}},
  {Kepler Data Processing Handbook: Transiting Planet Search}, Kepler Science
  Document KSCI-19081-003

\bibitem[{{Jenkins} {et~al.}(2010){Jenkins}, {Chandrasekaran}, {McCauliff},
  {Caldwell}, {Tenenbaum}, {Li}, {Klaus}, {Cote}, \& {Middour}}]{Jenkins:2010}
{Jenkins}, J.~M., {Chandrasekaran}, H., {McCauliff}, S.~D., {et~al.} 2010, in
  Society of Photo-Optical Instrumentation Engineers (SPIE) Conference Series,
  Vol. 7740, Software and Cyberinfrastructure for Astronomy, ed. N.~M.
  {Radziwill} \& A.~{Bridger}, 77400D, \dodoi{10.1117/12.856764}

\bibitem[{{Jenkins} {et~al.}(2016){Jenkins}, {Twicken}, {McCauliff},
  {Campbell}, {Sanderfer}, {Lung}, {Mansouri-Samani}, {Girouard}, {Tenenbaum},
  {Klaus}, {Smith}, {Caldwell}, {Chacon}, {Henze}, {Heiges}, {Latham},
  {Morgan}, {Swade}, {Rinehart}, \& {Vanderspek}}]{Jenkins:2016}
{Jenkins}, J.~M., {Twicken}, J.~D., {McCauliff}, S., {et~al.} 2016, in
  \procspie, Vol. 9913, Software and Cyberinfrastructure for Astronomy IV,
  99133E, \dodoi{10.1117/12.2233418}

\bibitem[{{Jenkins} {et~al.}(2020{\natexlab{b}}){Jenkins}, {D{\'\i}az},
  {Kurtovic}, {Espinoza}, {Vines}, {Rojas}, {Brahm}, {Torres},
  {Cort{\'e}s-Zuleta}, {Soto}, {Lopez}, {King}, {Wheatley}, {Winn}, {Ciardi},
  {Ricker}, {Vanderspek}, {Latham}, {Seager}, {Jenkins}, {Beichman}, {Bieryla},
  {Burke}, {Christiansen}, {Henze}, {Klaus}, {McCauliff}, {Mori}, {Narita},
  {Nishiumi}, {Tamura}, {de Leon}, {Quinn}, {Villase{\~n}or}, {Vezie},
  {Lissauer}, {Collins}, {Collins}, {Isopi}, {Mallia}, {Ercolino}, {Petrovich},
  {Jord{\'a}n}, {Acton}, {Armstrong}, {Bayliss}, {Bouchy}, {Belardi}, {Bryant},
  {Burleigh}, {Cabrera}, {Casewell}, {Chaushev}, {Cooke}, {Eigm{\"u}ller},
  {Erikson}, {Foxell}, {G{\"a}nsicke}, {Gill}, {Gillen}, {G{\"u}nther}, {Goad},
  {Hooton}, {Jackman}, {Louden}, {McCormac}, {Moyano}, {Nielsen}, {Pollacco},
  {Queloz}, {Rauer}, {Raynard}, {Smith}, {Tilbrook}, {Titz-Weider}, {Turner},
  {Udry}, {Walker}, {Watson}, {West}, {Palle}, {Ziegler}, {Law}, \&
  {Mann}}]{Jenkins:2020}
{Jenkins}, J.~S., {D{\'\i}az}, M.~R., {Kurtovic}, N.~T., {et~al.}
  2020{\natexlab{b}}, Nature Astronomy, 4, 1148,
  \dodoi{10.1038/s41550-020-1142-z}

\bibitem[{{Johnson} \& {Morgan}(1953)}]{Johnson:1953}
{Johnson}, H.~L., \& {Morgan}, W.~W. 1953, \apj, 117, 313,
  \dodoi{10.1086/145697}

\bibitem[{{Johnstone} {et~al.}(2021){Johnstone}, {Bartel}, \&
  {G{\"u}del}}]{Johnstone:2021}
{Johnstone}, C.~P., {Bartel}, M., \& {G{\"u}del}, M. 2021, \aap, 649, A96,
  \dodoi{10.1051/0004-6361/202038407}

\bibitem[{{Kass} \& {Raftery}(1995)}]{KassRaftery1995}
{Kass}, R., \& {Raftery}, A. 1995, Journal of the American Statistical
  Association, 90, 773

\bibitem[{{Kempton} {et~al.}(2018){Kempton}, {Bean}, {Louie}, {Deming}, {Koll},
  {Mansfield}, {Christiansen}, {L{\'o}pez-Morales}, {Swain}, {Zellem},
  {Ballard}, {Barclay}, {Barstow}, {Batalha}, {Beatty}, {Berta-Thompson},
  {Birkby}, {Buchhave}, {Charbonneau}, {Cowan}, {Crossfield}, {de Val-Borro},
  {Doyon}, {Dragomir}, {Gaidos}, {Heng}, {Hu}, {Kane}, {Kreidberg}, {Mallonn},
  {Morley}, {Narita}, {Nascimbeni}, {Pall{\'e}}, {Quintana}, {Rauscher},
  {Seager}, {Shkolnik}, {Sing}, {Sozzetti}, {Stassun}, {Valenti}, \& {von
  Essen}}]{Kempton:2018}
{Kempton}, E.~M.-R., {Bean}, J.~L., {Louie}, D.~R., {et~al.} 2018, \pasp, 130,
  114401, \dodoi{10.1088/1538-3873/aadf6f}

\bibitem[{{Kempton} {et~al.}(2023){Kempton}, {Zhang}, {Bean}, {Steinrueck},
  {Piette}, {Parmentier}, {Malsky}, {Roman}, {Rauscher}, {Gao}, {Bell}, {Xue},
  {Taylor}, {Savel}, {Arnold}, {Nixon}, {Stevenson}, {Mansfield}, {Kendrew},
  {Zieba}, {Ducrot}, {Dyrek}, {Lagage}, {Stassun}, {Henry}, {Barman}, {Lupu},
  {Malik}, {Kataria}, {Ih}, {Fu}, {Welbanks}, \& {McGill}}]{Kempton:2023}
{Kempton}, E. M.~R., {Zhang}, M., {Bean}, J.~L., {et~al.} 2023, \nat, 620, 67,
  \dodoi{10.1038/s41586-023-06159-5}

\bibitem[{{Kipping}(2013)}]{Kipping:2013}
{Kipping}, D.~M. 2013, \mnras, 435, 2152, \dodoi{10.1093/mnras/stt1435}

\bibitem[{{Kreidberg}(2015)}]{batman}
{Kreidberg}, L. 2015, \pasp, 127, 1161, \dodoi{10.1086/683602}

\bibitem[{{Kubyshkina} {et~al.}(2018){Kubyshkina}, {Fossati}, {Erkaev},
  {Cubillos}, {Johnstone}, {Kislyakova}, {Lammer}, {Lendl}, \&
  {Odert}}]{Kubyshkina:2018}
{Kubyshkina}, D., {Fossati}, L., {Erkaev}, N.~V., {et~al.} 2018, \apjl, 866,
  L18, \dodoi{10.3847/2041-8213/aae586}

\bibitem[{{Kunimoto} {et~al.}(2022{\natexlab{a}}){Kunimoto}, {Daylan},
  {Guerrero}, {Fong}, {Bryson}, {Ricker}, {Fausnaugh}, {Huang}, {Sha},
  {Shporer}, {Vanderburg}, {Vanderspek}, \& {Yu}}]{Kunimoto2022}
{Kunimoto}, M., {Daylan}, T., {Guerrero}, N., {et~al.} 2022{\natexlab{a}},
  \apjs, 259, 33, \dodoi{10.3847/1538-4365/ac5688}

\bibitem[{{Kunimoto} {et~al.}(2022{\natexlab{b}}){Kunimoto}, {Daylan},
  {Guerrero}, {Fong}, {Bryson}, {Ricker}, {Fausnaugh}, {Huang}, {Sha},
  {Shporer}, {Vanderburg}, {Vanderspek}, \& {Yu}}]{Kunimoto:2022}
---. 2022{\natexlab{b}}, \apjs, 259, 33, \dodoi{10.3847/1538-4365/ac5688}

\bibitem[{{Kurucz}(1993)}]{KuruczModel}
{Kurucz}, R. 1993, ATLAS9 Stellar Atmosphere Programs and 2 km/s grid. Kurucz
  CD-ROM No. 13. Cambridge, 13

\bibitem[{{Lallement} {et~al.}(2019){Lallement}, {Babusiaux}, {Vergely},
  {Katz}, {Arenou}, {Valette}, {Hottier}, \& {Capitanio}}]{Lallement2019}
{Lallement}, R., {Babusiaux}, C., {Vergely}, J.~L., {et~al.} 2019, \aap, 625,
  A135, \dodoi{10.1051/0004-6361/201834695}

\bibitem[{{Li} {et~al.}(2019){Li}, {Tenenbaum}, {Twicken}, {Burke}, {Jenkins},
  {Quintana}, {Rowe}, \& {Seader}}]{Li:2019}
{Li}, J., {Tenenbaum}, P., {Twicken}, J.~D., {et~al.} 2019, \pasp, 131, 024506,
  \dodoi{10.1088/1538-3873/aaf44d}

\bibitem[{{Lindegren} {et~al.}(2021){Lindegren}, {Bastian}, {Biermann},
  {Bombrun}, {de Torres}, {Gerlach}, {Geyer}, {Hern{\'a}ndez}, {Hilger},
  {Hobbs}, {Klioner}, {Lammers}, {McMillan}, {Ramos-Lerate},
  {Steidelm{\"u}ller}, {Stephenson}, \& {van Leeuwen}}]{DR3ParallaxCorrection}
{Lindegren}, L., {Bastian}, U., {Biermann}, M., {et~al.} 2021, \aap, 649, A4,
  \dodoi{10.1051/0004-6361/202039653}

\bibitem[{{Liu} {et~al.}(2013){Liu}, {Guillochon}, {Lin}, \&
  {Ramirez-Ruiz}}]{Liu:2013}
{Liu}, S.-F., {Guillochon}, J., {Lin}, D. N.~C., \& {Ramirez-Ruiz}, E. 2013,
  \apj, 762, 37, \dodoi{10.1088/0004-637X/762/1/37}

\bibitem[{{Lopez} \& {Fortney}(2013)}]{LopezFortney:2013}
{Lopez}, E.~D., \& {Fortney}, J.~J. 2013, \apj, 776, 2,
  \dodoi{10.1088/0004-637X/776/1/2}

\bibitem[{{Lopez} \& {Fortney}(2014)}]{LopezFortney:2014}
---. 2014, \apj, 792, 1, \dodoi{10.1088/0004-637X/792/1/1}

\bibitem[{{Lorenzo-Oliveira} {et~al.}(2016){Lorenzo-Oliveira}, {Porto de
  Mello}, \& {Schiavon}}]{Lorenzo-Oliveira:2016}
{Lorenzo-Oliveira}, D., {Porto de Mello}, G.~F., \& {Schiavon}, R.~P. 2016,
  \aap, 594, L3, \dodoi{10.1051/0004-6361/201629233}

\bibitem[{{McCall}(2004)}]{McCall2004}
{McCall}, M.~L. 2004, \aj, 128, 2144, \dodoi{10.1086/424933}

\bibitem[{{McCully} {et~al.}(2018){McCully}, {Volgenau}, {Harbeck}, {Lister},
  {Saunders}, {Turner}, {Siiverd}, \& {Bowman}}]{McCully:2018}
{McCully}, C., {Volgenau}, N.~H., {Harbeck}, D.-R., {et~al.} 2018, in Society
  of Photo-Optical Instrumentation Engineers (SPIE) Conference Series, Vol.
  10707, Software and Cyberinfrastructure for Astronomy V, ed. J.~C. {Guzman}
  \& J.~{Ibsen}, 107070K, \dodoi{10.1117/12.2314340}

\bibitem[{McKinney(2010)}]{McKinney2010}
McKinney, W. 2010, in Proceedings of the 9th Python in Science Conference, ed.
  S.~van~der Walt \& J.~Millman, 51 -- 56

\bibitem[{{Morton}(2015)}]{Isochrones}
{Morton}, T.~D. 2015, {isochrones: Stellar model grid package}, Astrophysics
  Source Code Library, record ascl:1503.010.
\newblock \doeprint{1503.010}

\bibitem[{{Naponiello} {et~al.}(2023){Naponiello}, {Mancini}, {Sozzetti},
  {Bonomo}, {Morbidelli}, {Dou}, {Zeng}, {Leinhardt}, {Biazzo}, {Cubillos},
  {Pinamonti}, {Locci}, {Maggio}, {Damasso}, {Lanza}, {Lissauer}, {Collins},
  {Carter}, {Jensen}, {Bignamini}, {Boschin}, {Bouma}, {Ciardi}, {Cosentino},
  {Crossfield}, {Desidera}, {Dumusque}, {Fiorenzano}, {Fukui}, {Giacobbe},
  {Gnilka}, {Ghedina}, {Guilluy}, {Harutyunyan}, {Howell}, {Jenkins}, {Lund},
  {Kielkopf}, {Lester}, {Malavolta}, {Mann}, {Matson}, {Matthews}, {Nardiello},
  {Narita}, {Pace}, {Pagano}, {Palle}, {Pedani}, {Seager}, {Schlieder},
  {Schwarz}, {Shporer}, {Twicken}, {Winn}, {Ziegler}, \&
  {Zingales}}]{Naponiello:2023}
{Naponiello}, L., {Mancini}, L., {Sozzetti}, A., {et~al.} 2023, \nat, 622, 255,
  \dodoi{10.1038/s41586-023-06499-2}

\bibitem[{{Onken} {et~al.}(2019){Onken}, {Wolf}, {Bessell}, {Chang}, {Da
  Costa}, {Luvaul}, {Mackey}, {Schmidt}, \& {Shao}}]{Onken:2019}
{Onken}, C.~A., {Wolf}, C., {Bessell}, M.~S., {et~al.} 2019, \pasa, 36, e033,
  \dodoi{10.1017/pasa.2019.27}

\bibitem[{{Osborn} {et~al.}(2023){Osborn}, {Armstrong}, {Fern{\'a}ndez
  Fern{\'a}ndez}, {Knierim}, {Adibekyan}, {Collins}, {Delgado-Mena},
  {Fridlund}, {Gomes da Silva}, {Hellier}, {Jackson}, {King}, {Lillo-Box},
  {Matson}, {Matthews}, {Santos}, {Sousa}, {Stassun}, {Tan}, {Ricker},
  {Vanderspek}, {Latham}, {Seager}, {Winn}, {Jenkins}, {Bayliss}, {Bouma},
  {Ciardi}, {Collins}, {Col{\'o}n}, {Crossfield}, {Demangeon}, {D{\'\i}az},
  {Dorn}, {Dumusque}, {Keniger}, {Figueira}, {Gan}, {Goeke}, {Hadjigeorghiou},
  {Hawthorn}, {Helled}, {Howell}, {Nielsen}, {Osborn}, {Quinn}, {Sefako},
  {Shporer}, {Str{\o}m}, {Twicken}, {Vanderburg}, \& {Wheatley}}]{Osborn:2023}
{Osborn}, A., {Armstrong}, D.~J., {Fern{\'a}ndez Fern{\'a}ndez}, J., {et~al.}
  2023, \mnras, 526, 548, \dodoi{10.1093/mnras/stad2575}

\bibitem[{{Owen} \& {Wu}(2013)}]{OwenWu2013}
{Owen}, J.~E., \& {Wu}, Y. 2013, \apj, 775, 105,
  \dodoi{10.1088/0004-637X/775/2/105}

\bibitem[{{Paegert} {et~al.}(2021){Paegert}, {Stassun}, {Collins}, {Pepper},
  {Torres}, {Jenkins}, {Twicken}, \& {Latham}}]{Paegert:2021}
{Paegert}, M., {Stassun}, K.~G., {Collins}, K.~A., {et~al.} 2021, arXiv
  e-prints, arXiv:2108.04778, \dodoi{10.48550/arXiv.2108.04778}

\bibitem[{{Papaloizou} \& {Terquem}(2010)}]{Papaloizou:2010}
{Papaloizou}, J. C.~B., \& {Terquem}, C. 2010, \mnras, 405, 573,
  \dodoi{10.1111/j.1365-2966.2010.16477.x}

\bibitem[{{Pepe} {et~al.}(2000){Pepe}, {Mayor}, {Delabre}, {Kohler}, {Lacroix},
  {Queloz}, {Udry}, {Benz}, {Bertaux}, \& {Sivan}}]{Pepe:2000}
{Pepe}, F., {Mayor}, M., {Delabre}, B., {et~al.} 2000, in Society of
  Photo-Optical Instrumentation Engineers (SPIE) Conference Series, Vol. 4008,
  Optical and IR Telescope Instrumentation and Detectors, ed. M.~{Iye} \& A.~F.
  {Moorwood}, 582--592, \dodoi{10.1117/12.395516}

\bibitem[{{Pepe} {et~al.}(2002){Pepe}, {Mayor}, {Rupprecht}, {Avila},
  {Ballester}, {Beckers}, {Benz}, {Bertaux}, {Bouchy}, {Buzzoni}, {Cavadore},
  {Deiries}, {Dekker}, {Delabre}, {D'Odorico}, {Eckert}, {Fischer}, {Fleury},
  {George}, {Gilliotte}, {Gojak}, {Guzman}, {Koch}, {Kohler}, {Kotzlowski},
  {Lacroix}, {Le Merrer}, {Lizon}, {Lo Curto}, {Longinotti}, {Megevand},
  {Pasquini}, {Petitpas}, {Pichard}, {Queloz}, {Reyes}, {Richaud}, {Sivan},
  {Sosnowska}, {Soto}, {Udry}, {Ureta}, {van Kesteren}, {Weber}, {Weilenmann},
  {Wicenec}, {Wieland}, {Christensen-Dalsgaard}, {Dravins}, {Hatzes},
  {K{\"u}rster}, {Paresce}, \& {Penny}}]{Pepe2002}
{Pepe}, F., {Mayor}, M., {Rupprecht}, G., {et~al.} 2002, The Messenger, 110, 9

\bibitem[{{Pepe} {et~al.}(2021){Pepe}, {Cristiani}, {Rebolo}, {Santos},
  {Dekker}, {Cabral}, {Di Marcantonio}, {Figueira}, {Lo Curto}, {Lovis},
  {Mayor}, {M{\'e}gevand}, {Molaro}, {Riva}, {Zapatero Osorio}, {Amate},
  {Manescau}, {Pasquini}, {Zerbi}, {Adibekyan}, {Abreu}, {Affolter}, {Alibert},
  {Aliverti}, {Allart}, {Allende Prieto}, {{\'A}lvarez}, {Alves}, {Avila},
  {Baldini}, {Bandy}, {Barros}, {Benz}, {Bianco}, {Borsa}, {Bourrier},
  {Bouchy}, {Broeg}, {Calderone}, {Cirami}, {Coelho}, {Conconi}, {Coretti},
  {Cumani}, {Cupani}, {D'Odorico}, {Damasso}, {Deiries}, {Delabre},
  {Demangeon}, {Dumusque}, {Ehrenreich}, {Faria}, {Fragoso}, {Genolet},
  {Genoni}, {G{\'e}nova Santos}, {Gonz{\'a}lez Hern{\'a}ndez}, {Hughes},
  {Iwert}, {Kerber}, {Knudstrup}, {Landoni}, {Lavie}, {Lillo-Box}, {Lizon},
  {Maire}, {Martins}, {Mehner}, {Micela}, {Modigliani}, {Monteiro}, {Monteiro},
  {Moschetti}, {Murphy}, {Nunes}, {Oggioni}, {Oliveira}, {Oshagh}, {Pall{\'e}},
  {Pariani}, {Poretti}, {Rasilla}, {Rebord{\~a}o}, {Redaelli}, {Santana
  Tschudi}, {Santin}, {Santos}, {S{\'e}gransan}, {Schmidt}, {Segovia},
  {Sosnowska}, {Sozzetti}, {Sousa}, {Span{\`o}}, {Su{\'a}rez Mascare{\~n}o},
  {Tabernero}, {Tenegi}, {Udry}, \& {Zanutta}}]{Pepe2021}
{Pepe}, F., {Cristiani}, S., {Rebolo}, R., {et~al.} 2021, \aap, 645, A96,
  \dodoi{10.1051/0004-6361/202038306}

\bibitem[{{Qin} {et~al.}(2023){Qin}, {Zhong}, {Tang}, \& {Chen}}]{Qin2023}
{Qin}, S., {Zhong}, J., {Tang}, T., \& {Chen}, L. 2023, \apjs, 265, 12,
  \dodoi{10.3847/1538-4365/acadd6}

\bibitem[{{Saar} \& {Donahue}(1997)}]{SaarDonahue1997}
{Saar}, S.~H., \& {Donahue}, R.~A. 1997, \apj, 485, 319, \dodoi{10.1086/304392}

\bibitem[{{Santos} {et~al.}(2013){Santos}, {Sousa}, {Mortier}, {Neves},
  {Adibekyan}, {Tsantaki}, {Delgado Mena}, {Bonfils}, {Israelian}, {Mayor}, \&
  {Udry}}]{Santos-13}
{Santos}, N.~C., {Sousa}, S.~G., {Mortier}, A., {et~al.} 2013, \aap, 556, A150,
  \dodoi{10.1051/0004-6361/201321286}

\bibitem[{{Schlegel} {et~al.}(1998){Schlegel}, {Finkbeiner}, \&
  {Davis}}]{Schlegel1998}
{Schlegel}, D.~J., {Finkbeiner}, D.~P., \& {Davis}, M. 1998, \apj, 500, 525,
  \dodoi{10.1086/305772}

\bibitem[{{Shallue} \& {Vanderburg}(2018)}]{ShallueVanderburg2018}
{Shallue}, C.~J., \& {Vanderburg}, A. 2018, \aj, 155, 94,
  \dodoi{10.3847/1538-3881/aa9e09}

\bibitem[{{Skrutskie} {et~al.}(2006){Skrutskie}, {Cutri}, {Stiening},
  {Weinberg}, {Schneider}, {Carpenter}, {Beichman}, {Capps}, {Chester},
  {Elias}, {Huchra}, {Liebert}, {Lonsdale}, {Monet}, {Price}, {Seitzer},
  {Jarrett}, {Kirkpatrick}, {Gizis}, {Howard}, {Evans}, {Fowler}, {Fullmer},
  {Hurt}, {Light}, {Kopan}, {Marsh}, {McCallon}, {Tam}, {Van Dyk}, \&
  {Wheelock}}]{Skrutskie2006}
{Skrutskie}, M.~F., {Cutri}, R.~M., {Stiening}, R., {et~al.} 2006, \aj, 131,
  1163

\bibitem[{{Smith} {et~al.}(2012){Smith}, {Stumpe}, {Van Cleve}, {Jenkins},
  {Barclay}, {Fanelli}, {Girouard}, {Kolodziejczak}, {McCauliff}, {Morris}, \&
  {Twicken}}]{Smith:2012}
{Smith}, J.~C., {Stumpe}, M.~C., {Van Cleve}, J.~E., {et~al.} 2012, \pasp, 124,
  1000, \dodoi{10.1086/667697}

\bibitem[{{Sneden}(1973)}]{Sneden-73}
{Sneden}, C.~A. 1973, PhD thesis, THE UNIVERSITY OF TEXAS AT AUSTIN.

\bibitem[{{Soto} \& {Jenkins}(2018)}]{species}
{Soto}, M.~G., \& {Jenkins}, J.~S. 2018, \aap, 615, A76,
  \dodoi{10.1051/0004-6361/201731533}

\bibitem[{{Soto} {et~al.}(2021){Soto}, {Jones}, \& {Jenkins}}]{Soto:2021}
{Soto}, M.~G., {Jones}, M.~I., \& {Jenkins}, J.~S. 2021, \aap, 647, A157,
  \dodoi{10.1051/0004-6361/202039357}

\bibitem[{{Sousa}(2014)}]{Sousa-14}
{Sousa}, S.~G. 2014, [arXiv:1407.5817].
\newblock \doarXiv{1407.5817}

\bibitem[{{Sousa} {et~al.}(2015){Sousa}, {Santos}, {Adibekyan}, {Delgado-Mena},
  \& {Israelian}}]{Sousa-15}
{Sousa}, S.~G., {Santos}, N.~C., {Adibekyan}, V., {Delgado-Mena}, E., \&
  {Israelian}, G. 2015, \aap, 577, A67, \dodoi{10.1051/0004-6361/201425463}

\bibitem[{{Sousa} {et~al.}(2007){Sousa}, {Santos}, {Israelian}, {Mayor}, \&
  {Monteiro}}]{Sousa-07}
{Sousa}, S.~G., {Santos}, N.~C., {Israelian}, G., {Mayor}, M., \& {Monteiro},
  M.~J.~P.~F.~G. 2007, A\&A, 469, 783, \dodoi{10.1051/0004-6361:20077288}

\bibitem[{{Sousa} {et~al.}(2021){Sousa}, {Adibekyan}, {Delgado-Mena}, {Santos},
  {Rojas-Ayala}, {Soares}, {Legoinha}, {Ulmer-Moll}, {Camacho}, {Barros},
  {Demangeon}, {Hoyer}, {Israelian}, {Mortier}, {Tsantaki}, \&
  {Monteiro}}]{Sousa-21}
{Sousa}, S.~G., {Adibekyan}, V., {Delgado-Mena}, E., {et~al.} 2021, arXiv
  e-prints, arXiv:2109.04781.
\newblock \doarXiv{2109.04781}

\bibitem[{{Spake} {et~al.}(2018){Spake}, {Sing}, {Evans}, {Oklop{\v c}i{\'c}},
  {Bourrier}, {Kreidberg}, {Rackham}, {Irwin}, {Ehrenreich}, {Wyttenbach},
  {Wakeford}, {Zhou}, {Chubb}, {Nikolov}, {Goyal}, {Henry}, {Williamson},
  {Blumenthal}, {Anderson}, {Hellier}, {Charbonneau}, {Udry}, \&
  {Madhusudhan}}]{Spake:2018}
{Spake}, J.~J., {Sing}, D.~K., {Evans}, T.~M., {et~al.} 2018, \nat, 557, 68,
  \dodoi{10.1038/s41586-018-0067-5}

\bibitem[{{Stassun} {et~al.}(2019{\natexlab{a}}){Stassun}, {Oelkers},
  {Paegert}, {Torres}, {Pepper}, {De Lee}, {Collins}, {Latham}, {Muirhead},
  {Chittidi}, {Rojas-Ayala}, {Fleming}, {Rose}, {Tenenbaum}, {Ting}, {Kane},
  {Barclay}, {Bean}, {Brassuer}, {Charbonneau}, {Ge}, {Lissauer}, {Mann},
  {McLean}, {Mullally}, {Narita}, {Plavchan}, {Ricker}, {Sasselov}, {Seager},
  {Sharma}, {Shiao}, {Sozzetti}, {Stello}, {Vanderspek}, {Wallace}, \&
  {Winn}}]{Stassun:2019}
{Stassun}, K.~G., {Oelkers}, R.~J., {Paegert}, M., {et~al.} 2019{\natexlab{a}},
  \aj, 158, 138, \dodoi{10.3847/1538-3881/ab3467}

\bibitem[{{Stassun} {et~al.}(2019{\natexlab{b}}){Stassun}, {Oelkers},
  {Paegert}, {Torres}, {Pepper}, {De Lee}, {Collins}, {Latham}, {Muirhead},
  {Chittidi}, {Rojas-Ayala}, {Fleming}, {Rose}, {Tenenbaum}, {Ting}, {Kane},
  {Barclay}, {Bean}, {Brassuer}, {Charbonneau}, {Ge}, {Lissauer}, {Mann},
  {McLean}, {Mullally}, {Narita}, {Plavchan}, {Ricker}, {Sasselov}, {Seager},
  {Sharma}, {Shiao}, {Sozzetti}, {Stello}, {Vanderspek}, {Wallace}, \&
  {Winn}}]{Stassun2019}
---. 2019{\natexlab{b}}, \aj, 158, 138, \dodoi{10.3847/1538-3881/ab3467}

\bibitem[{{Stumpe} {et~al.}(2014){Stumpe}, {Smith}, {Catanzarite}, {Van Cleve},
  {Jenkins}, {Twicken}, \& {Girouard}}]{Stumpe:2014}
{Stumpe}, M.~C., {Smith}, J.~C., {Catanzarite}, J.~H., {et~al.} 2014, \pasp,
  126, 100, \dodoi{10.1086/674989}

\bibitem[{{Stumpe} {et~al.}(2012){Stumpe}, {Smith}, {Van Cleve}, {Twicken},
  {Barclay}, {Fanelli}, {Girouard}, {Jenkins}, {Kolodziejczak}, {McCauliff}, \&
  {Morris}}]{Stumpe:2012}
{Stumpe}, M.~C., {Smith}, J.~C., {Van Cleve}, J.~E., {et~al.} 2012, \pasp, 124,
  985, \dodoi{10.1086/667698}

\bibitem[{{Szab{\'o}} \& {Kiss}(2011)}]{NeptuneDesert}
{Szab{\'o}}, G.~M., \& {Kiss}, L.~L. 2011, \apjl, 727, L44,
  \dodoi{10.1088/2041-8205/727/2/L44}

\bibitem[{{Thorngren} {et~al.}(2023){Thorngren}, {Lee}, \&
  {Lopez}}]{Thorngren:2023}
{Thorngren}, D.~P., {Lee}, E.~J., \& {Lopez}, E.~D. 2023, \apjl, 945, L36,
  \dodoi{10.3847/2041-8213/acbd35}

\bibitem[{{Tokovinin} \& {Cantarutti}(2008)}]{SoarSpeckle}
{Tokovinin}, A., \& {Cantarutti}, R. 2008, \pasp, 120, 170,
  \dodoi{10.1086/528809}

\bibitem[{{Tokovinin} {et~al.}(2022){Tokovinin}, {Mason}, {Mendez}, \&
  {Costa}}]{Tokovinin:2022}
{Tokovinin}, A., {Mason}, B.~D., {Mendez}, R.~A., \& {Costa}, E. 2022, \aj,
  164, 58, \dodoi{10.3847/1538-3881/ac78e7}

\bibitem[{{Triaud}(2017)}]{Triaud2017}
{Triaud}, A.~H.~M.~J. 2017, {The Rossiter-McLaughlin Effect in Exoplanet
  Research} (Springer, Cham), 2, \dodoi{10.1007/978-3-319-30648-3_2-1}

\bibitem[{{Tsantaki} {et~al.}(2013){Tsantaki}, {Sousa}, {Adibekyan}, {Santos},
  {Mortier}, \& {Israelian}}]{Tsantaki-2013}
{Tsantaki}, M., {Sousa}, S.~G., {Adibekyan}, V.~Z., {et~al.} 2013, \aap, 555,
  A150, \dodoi{10.1051/0004-6361/201321103}

\bibitem[{{Twicken} {et~al.}(2018){Twicken}, {Catanzarite}, {Clarke},
  {Girouard}, {Jenkins}, {Klaus}, {Li}, {McCauliff}, {Seader}, {Tenenbaum},
  {Wohler}, {Bryson}, {Burke}, {Caldwell}, {Haas}, {Henze}, \&
  {Sanderfer}}]{Twicken:2018}
{Twicken}, J.~D., {Catanzarite}, J.~H., {Clarke}, B.~D., {et~al.} 2018, \pasp,
  130, 064502, \dodoi{10.1088/1538-3873/aab694}

\bibitem[{{Valsecchi} {et~al.}(2015){Valsecchi}, {Rappaport}, {Rasio},
  {Marchant}, \& {Rogers}}]{Valsecchi:2015}
{Valsecchi}, F., {Rappaport}, S., {Rasio}, F.~A., {Marchant}, P., \& {Rogers},
  L.~A. 2015, \apj, 813, 101, \dodoi{10.1088/0004-637X/813/2/101}

\bibitem[{{Valsecchi} {et~al.}(2014){Valsecchi}, {Rasio}, \&
  {Steffen}}]{Valsecchi:2014}
{Valsecchi}, F., {Rasio}, F.~A., \& {Steffen}, J.~H. 2014, \apjl, 793, L3,
  \dodoi{10.1088/2041-8205/793/1/L3}

\bibitem[{{Vanderburg} \& {Johnson}(2014)}]{vj14}
{Vanderburg}, A., \& {Johnson}, J.~A. 2014, \pasp, 126, 948,
  \dodoi{10.1086/678764}

\bibitem[{{Vanderburg} {et~al.}(2019){Vanderburg}, {Huang}, {Rodriguez},
  {Becker}, {Ricker}, {Vanderspek}, {Latham}, {Seager}, {Winn}, {Jenkins},
  {Addison}, {Bieryla}, {Brice{\~n}o}, {Bowler}, {Brown}, {Burke}, {Burt},
  {Caldwell}, {Clark}, {Crossfield}, {Dittmann}, {Dynes}, {Fulton}, {Guerrero},
  {Harbeck}, {Horner}, {Kane}, {Kielkopf}, {Kraus}, {Kreidberg}, {Law}, {Mann},
  {Mengel}, {Morton}, {Okumura}, {Pearce}, {Plavchan}, {Quinn}, {Rabus},
  {Rose}, {Rowden}, {Shporer}, {Siverd}, {Smith}, {Stassun}, {Tinney},
  {Wittenmyer}, {Wright}, {Zhang}, {Zhou}, \& {Ziegler}}]{Vanderburg:2019}
{Vanderburg}, A., {Huang}, C.~X., {Rodriguez}, J.~E., {et~al.} 2019, \apjl,
  881, L19, \dodoi{10.3847/2041-8213/ab322d}

\bibitem[{{Vines} \& {Jenkins}(2022)}]{ariadne}
{Vines}, J.~I., \& {Jenkins}, J.~S. 2022, \mnras, 513, 2719,
  \dodoi{10.1093/mnras/stac956}

\bibitem[{{Virtanen} {et~al.}(2020){Virtanen}, {Gommers}, {Oliphant},
  {Haberland}, {Reddy}, {Cournapeau}, {Burovski}, {Peterson}, {Weckesser},
  {Bright}, {van der Walt}, {Brett}, {Wilson}, {Millman}, {Mayorov}, {Nelson},
  {Jones}, {Kern}, {Larson}, {Carey}, {Polat}, {Feng}, {Moore}, {VanderPlas},
  {Laxalde}, {Perktold}, {Cimrman}, {Henriksen}, {Quintero}, {Harris},
  {Archibald}, {Ribeiro}, {Pedregosa}, {van Mulbregt}, \& {SciPy 1. 0
  Contributors}}]{Virtanen:2020}
{Virtanen}, P., {Gommers}, R., {Oliphant}, T.~E., {et~al.} 2020, Nature
  Methods, 17, 261, \dodoi{10.1038/s41592-019-0686-2}

\bibitem[{{Winn} {et~al.}(2017){Winn}, {Sanchis-Ojeda}, {Rogers}, {Petigura},
  {Howard}, {Isaacson}, {Marcy}, {Schlaufman}, {Cargile}, \&
  {Hebb}}]{Winn:2017}
{Winn}, J.~N., {Sanchis-Ojeda}, R., {Rogers}, L., {et~al.} 2017, \aj, 154, 60,
  \dodoi{10.3847/1538-3881/aa7b7c}

\bibitem[{{Wright} {et~al.}(2010){Wright}, {Eisenhardt}, {Mainzer}, {Ressler},
  {Cutri}, {Jarrett}, {Kirkpatrick}, {Padgett}, {McMillan}, {Skrutskie},
  {Stanford}, {Cohen}, {Walker}, {Mather}, {Leisawitz}, {Gautier}, {McLean},
  {Benford}, {Lonsdale}, {Blain}, {Mendez}, {Irace}, {Duval}, {Liu}, {Royer},
  {Heinrichsen}, {Howard}, {Shannon}, {Kendall}, {Walsh}, {Larsen}, {Cardon},
  {Schick}, {Schwalm}, {Abid}, {Fabinsky}, {Naes}, \& {Tsai}}]{Wright2010}
{Wright}, E.~L., {Eisenhardt}, P.~R.~M., {Mainzer}, A.~K., {et~al.} 2010, \aj,
  140, 1868

\bibitem[{{York} {et~al.}(2000){York}, {Adelman}, {Anderson}, {Anderson},
  {Annis}, {Bahcall}, {Bakken}, {Barkhouser}, {Bastian}, {Berman}, {Boroski},
  {Bracker}, {Briegel}, {Briggs}, {Brinkmann}, {Brunner}, {Burles}, {Carey},
  {Carr}, {Castander}, {Chen}, {Colestock}, {Connolly}, {Crocker}, {Csabai},
  {Czarapata}, {Davis}, {Doi}, {Dombeck}, {Eisenstein}, {Ellman}, {Elms},
  {Evans}, {Fan}, {Federwitz}, {Fiscelli}, {Friedman}, {Frieman}, {Fukugita},
  {Gillespie}, {Gunn}, {Gurbani}, {de Haas}, {Haldeman}, {Harris}, {Hayes},
  {Heckman}, {Hennessy}, {Hindsley}, {Holm}, {Holmgren}, {Huang}, {Hull},
  {Husby}, {Ichikawa}, {Ichikawa}, {Ivezi{\'c}}, {Kent}, {Kim}, {Kinney},
  {Klaene}, {Kleinman}, {Kleinman}, {Knapp}, {Korienek}, {Kron}, {Kunszt},
  {Lamb}, {Lee}, {Leger}, {Limmongkol}, {Lindenmeyer}, {Long}, {Loomis},
  {Loveday}, {Lucinio}, {Lupton}, {MacKinnon}, {Mannery}, {Mantsch}, {Margon},
  {McGehee}, {McKay}, {Meiksin}, {Merelli}, {Monet}, {Munn}, {Narayanan},
  {Nash}, {Neilsen}, {Neswold}, {Newberg}, {Nichol}, {Nicinski}, {Nonino},
  {Okada}, {Okamura}, {Ostriker}, {Owen}, {Pauls}, {Peoples}, {Peterson},
  {Petravick}, {Pier}, {Pope}, {Pordes}, {Prosapio}, {Rechenmacher}, {Quinn},
  {Richards}, {Richmond}, {Rivetta}, {Rockosi}, {Ruthmansdorfer}, {Sandford},
  {Schlegel}, {Schneider}, {Sekiguchi}, {Sergey}, {Shimasaku}, {Siegmund},
  {Smee}, {Smith}, {Snedden}, {Stone}, {Stoughton}, {Strauss}, {Stubbs},
  {SubbaRao}, {Szalay}, {Szapudi}, {Szokoly}, {Thakar}, {Tremonti}, {Tucker},
  {Uomoto}, {Vanden Berk}, {Vogeley}, {Waddell}, {Wang}, {Watanabe},
  {Weinberg}, {Yanny}, {Yasuda}, \& {SDSS Collaboration}}]{York:2000}
{York}, D.~G., {Adelman}, J., {Anderson}, John~E., J., {et~al.} 2000, \aj, 120,
  1579, \dodoi{10.1086/301513}

\bibitem[{{Zacharias} {et~al.}(2013){Zacharias}, {Finch}, {Girard}, {Henden},
  {Bartlett}, {Monet}, \& {Zacharias}}]{Zacharias:2013}
{Zacharias}, N., {Finch}, C.~T., {Girard}, T.~M., {et~al.} 2013, \aj, 145, 44,
  \dodoi{10.1088/0004-6256/145/2/44}

\bibitem[{{Zellem} {et~al.}(2017){Zellem}, {Swain}, {Roudier}, {Shkolnik},
  {Creech-Eakman}, {Ciardi}, {Line}, {Iyer}, {Bryden}, {Llama}, \&
  {Fahy}}]{Zellem:2017}
{Zellem}, R.~T., {Swain}, M.~R., {Roudier}, G., {et~al.} 2017, \apj, 844, 27,
  \dodoi{10.3847/1538-4357/aa79f5}

\bibitem[{{Zeng} {et~al.}(2016){Zeng}, {Sasselov}, \& {Jacobsen}}]{Zeng:2016}
{Zeng}, L., {Sasselov}, D.~D., \& {Jacobsen}, S.~B. 2016, \apj, 819, 127,
  \dodoi{10.3847/0004-637X/819/2/127}

\bibitem[{{Ziegler} {et~al.}(2021){Ziegler}, {Tokovinin}, {Latiolais},
  {Brice{\~n}o}, {Law}, \& {Mann}}]{Ziegler:2021}
{Ziegler}, C., {Tokovinin}, A., {Latiolais}, M., {et~al.} 2021, \aj, 162, 192,
  \dodoi{10.3847/1538-3881/ac17f6}

\end{thebibliography}

\begin{deluxetable*}{lccr}

\tablewidth{0pc}
\tabletypesize{\scriptsize}
\tablecaption{
    Stellar and Planet Parameters for \target
    \label{tab:fit}
}
\tablehead{
    \multicolumn{1}{c}{~~~~~~~~Parameter~~~~~~~~} &
    \multicolumn{1}{c}{Prior}                     &
    \multicolumn{1}{c}{Fitted Value}              &
    \multicolumn{1}{c}{Source}    
}
\startdata
\noalign{\vskip -3pt}
\sidehead{Stellar Parameters}
~~~~$\mstar$ ($\msun$)\dotfill & U(0.7, 0.89) &  \starMass & \\
~~~~$\rstar$ ($\rsun$)\dotfill & U(0.82,0.98) & \starRadius & \\
~~~~$\loggstar$ (cgs)\dotfill & - & \starLogg &  ESPRESSO \\
~~~~$\lstar$ ($\lsun$)\dotfill & - & \starLuminosity  &  \\
~~~~$\teffstar$ (K)\dotfill & G(5065, 72) &  \starTeff  & \\
~~~~$\feh$ \dotfill & G(0.14, 0.05) &  \starfeh  & \\
~~~~Distance (pc)\dotfill   &  & \starDistance  & Parallax\\
~~~~\rhostar (\gcmc)\dotfill & $-$ &  \starRho &  \\
\sidehead{Limb-darkening coefficients}
~~~$q_1,TESS$               \dotfill & U(0, 1) & \starTESSqOne  &  \\
~~~$q_2,TESS$               \dotfill & U(0, 1) &  \starTESSqTwo & \\
~~~$q_1,LCO$               \dotfill & U(0, 1) & \starLCOqOne  &  \\
~~~$q_2,LCO$               \dotfill & U(0, 1) &  \starLCOqTwo  & \\
~~~$u_1,TESS$               \dotfill & $-$ & \starTESSuOne     &  \\
~~~$u_2,TESS$               \dotfill & $-$ &  \starTESSuTwo    & \\
~~~$u_1,LCO$               \dotfill  & $-$ & \starLCOuOne     &  \\
~~~$u_2,LCO$               \dotfill  & $-$ &  \starLCOuTwo    & \\
\sidehead{Light curve parameters}
~~~$P$ (days)  \dotfill    & U(0.883116, 0.883144) &  \bPeriod & \\
~~~$T_0$ \dotfill    & U(2459112.3988369, 2459112.4013631) &  \bEpoch \\
~~~$T_{14}$ (hr) \dotfill   & - &   \bDuration \\
~~~$T_{12} = T_{34}$ (min)   \dotfill    & $-$ &  \bIngressDuration \\
~~~$\arstar$              \dotfill & $-$ & \bAOR \\
~~~$\rpl/\rstar$          \dotfill & U(0, 1) & \bROR \\
~~~$b \equiv a \cos i/\rstar$ \dotfill   & U(0, 1) & \bImpactParameter \\
~~~$i$ (deg) \dotfill & $-$ &  \bInclination \\
\sidehead{Radial velocity parameters (m\,s$^{-1}$)}
~~~$K$          \dotfill & U(10, 30) & \bSemiAmplitude \\
~~~Jitter (ESPRESSO) \dotfill   & U(0, 20) & \jitterESPRESSO \\
~~~Jitter (HARPS) \dotfill   & U(0, 20) & \jitterHARPS \\
~~~$\gamma_\text{ESPRESSO}$ \dotfill & U(-2000, 1000) &  \gammaESPRESSO \\
~~~$\gamma_\text{HARPS}$ \dotfill & U(-2000, 1000) &  \gammaHARPS \\
\sidehead{Planetary parameters}
~~~${\rho}_p$ (g cm$^{-3}$) \dotfill & $-$ &  \bDensity \\
~~~$e$ \dotfill & $-$ &  \bEcc \tablenotemark{a}\\
~~~$\omega$ ($^\circ$) \dotfill & $-$ & \bomega \tablenotemark{a} \\
~~~$\mpl$ ($M_{\earth}$) \dotfill & $-$ &  \bMass \\
~~~$\rpl$ ($R_{\earth}$) \dotfill & $-$ &  \bRadius \\
~~~$a$ (AU) \dotfill & $-$ & \bSemimajorAxis \\
~~~$T_{\rm eq}$ (K) \dotfill & $-$ &   \bTeq \\
~~~$\langle F \rangle$ ($S_\oplus$) \dotfill  & $-$ & \bIrr \\
\enddata
\tablenotetext{a}{We only consider the case of a circular orbit and fix these parameters to be zero (see Section \ref{sec:rvmodel} for more details).}
\vspace{-.25cm}

\end{deluxetable*}

\begin{deluxetable*}{ccccccccc}
\tablewidth{0pt}
\tablecaption{
  Time-Series Radial Velocity Observations
  \label{tab:rv}
}
\tablehead{ 
    \colhead{Instrument} & \colhead{Date (UT)} & \colhead{Exposure Time (s)} & \colhead{SNR}  & \colhead{BJD} & \colhead{$RV_\text{CCF}$ (m\,s$^{-1}$)} & \colhead{$\sigma_\text{CCF}$ (m\,s$^{-1}$)}
}
\startdata
HARPS & 2023 Feb 3 & 2400 & 9.1 & 2459978.607357 & -1915 & 16\\
HARPS & 2023 Feb 5 & 2400 & 6.2 & 2459980.583111 & -1949 & 22\\	
HARPS & 2023 Feb 5 & 2400 & 7.2 & 2459980.646639 & -1931 & 19\\	
HARPS & 2023 Feb 6 & 2400 & 8.6 & 2459981.548598 & -1959 & 16\\
HARPS & 2023 Feb 6 & 2400 & 8.6 & 2459981.620540 & -1964 & 15\\	
HARPS & 2023 Feb 7 & 2400 & 12.7 & 2459982.548975 & -1941 & 9\\
HARPS & 2023 Feb 7 & 2400 & 9.9 & 2459982.609663 & -1956 & 13\\	
HARPS & 2023 Feb 8 & 2400 & 7.5 & 2459983.545592 & -1937 & 17\\	
HARPS & 2023 Feb 8 & 2400 & 7.3 & 2459983.613030 & -1913 & 18\\
HARPS & 2023 Mar 13 & 2400 & 7.1 & 2460016.541451 & -1918 & 18\\	
HARPS & 2023 Mar 14 & 2400 & 7.9 & 2460017.515069 & -1903 & 16\\	
HARPS & 2023 Mar 15 & 2400 & 10.7 & 2460018.519669 & -1930 & 11\\	
HARPS & 2023 Mar 16 & 2400 & 10.6 & 2460019.517703 & -1945 & 11\\	
HARPS & 2023 Mar 17 & 2400 & 11.5 & 2460020.516404 & -1911 & 10\\	
HARPS & 2023 Mar 18 & 2400 & 12.5 & 2460021.515254 & -1943.1 & 8.7\\
HARPS & 2023 Mar 19 & 2400 & 10.4 & 2460022.514002 & -1923 & 12\\	
HARPS & 2023 Jun 29 & 2400 & 4.6 & 2460124.890790 & -1918 & 32\\
HARPS & 2023 Jul 1 & 2400 & 10.9 & 2460126.909254 & -1902 & 11\\
HARPS & 2023 Jul 2 & 2400 & 9.4 & 2460127.937958 & -1893 & 14\\
HARPS & 2023 Jul 3 & 2400 & 9.0 & 2460128.927334 & -1947 & 16\\
HARPS & 2023 Jul 4 & 2400 & 8.6 & 2460129.901670 & -1963 & 16\\
HARPS & 2023 Jul 5 & 2400 & 8.8 & 2460130.891342 & -1935 & 15\\
HARPS & 2023 Jul 8 & 2400 & 8.6 & 2460133.937007 & -1899 & 15\\
HARPS & 2023 Jul 9 & 629.22\tablenotemark{a} & 0.5 & 2460134.924410 & -1900 & 200\\	
HARPS & 2023 Jul 14 & 2400 & 10.4 & 2460139.885208 & -1938 & 12\\
HARPS & 2023 Jul 15 & 2400 & 13.0 & 2460140.891320 & -1903.6 & 9.0\\
HARPS & 2023 Jul 16 & 2400 & 9.9 & 2460141.856275 & -1880 & 12\\	
HARPS & 2023 Jul 28 & 2400 & 11.3 & 2460153.819629 & -1943 & 11\\
HARPS & 2023 Jul 29 & 2400 & 11.9 & 2460154.861392 & -1946 & 10\\
HARPS & 2023 Jul 30 & 2400 & 9.9 & 2460155.827033 & -1931 & 13\\
HARPS & 2023 Jul 31 & 2400 & 12.2 & 2460156.883407 & -1907 & 10\\
HARPS & 2023 Aug 2 & 2400 & 14.5 & 2460158.818738 & -1900.0 & 8.6\\
HARPS & 2023 Aug 4 & 2400 & 12.9 & 2460160.811975 & -1935.1 & 9.7\\	
HARPS & 2023 Aug 20 & 2400 & 7.3 & 2460176.810481 & -1922 & 18\\	
HARPS & 2023 Aug 22 & 2400 & 7.0 & 2460178.822883 & -1913 & 19\\
HARPS & 2023 Aug 23 & 2400 & 10.1 & 2460179.869876 & -1887 & 12\\
ESPRESSO & 2023 Sep 19 & 1200 & 36.5 & 2460207.731305 & -1932.1 & 1.3\\
ESPRESSO & 2023 Sep 19 & 1200 & 38.7 & 2460207.748374 & -1931.4 & 1.3\\
ESPRESSO & 2023 Sep 19 & 1200 & 34.9 & 2460207.851007 & -1923.7 & 1.4\\
ESPRESSO & 2023 Sep 19 & 1200 & 34.8 & 2460207.868212 & -1918.9 & 1.4\\
ESPRESSO & 2023 Sep 20 & 1000 & 25.2 & 2460208.763323 & -1916.2 & 2.1\\
ESPRESSO & 2023 Sep 20 & 1000 & 20.6 & 2460208.808343 & -1899.2 & 2.7\\
ESPRESSO & 2023 Sep 20 & 1200 & 22.2 & 2460208.870812 & -1899.4 & 2.5\\
ESPRESSO & 2023 Sep 21 & 1200 & 29.1 & 2460209.741648 & -1902.3 & 1.8\\
ESPRESSO & 2023 Sep 21 & 1200 & 29.3 & 2460209.819560 & -1888.5 & 1.8\\
ESPRESSO & 2023 Sep 21 & 1200 & 28.9 & 2460209.835739 & -1888.1 & 1.8\\
ESPRESSO & 2023 Sep 21 & 1200 & 26.9 & 2460209.872248 & -1887.4 & 2.0\\
ESPRESSO & 2023 Sep 22 & 1200 & 35.5 & 2460210.738513 & -1892.4 & 1.4\\
ESPRESSO & 2023 Sep 22 & 1200 & 25.5 & 2460210.776400 & -1890.4 & 2.1\\
ESPRESSO & 2023 Sep 22 & 1200 & 23.2 & 2460210.848955 & -1893.4 & 2.3\\
ESPRESSO & 2023 Sep 22 & 1200 & 27.1 & 2460210.880491 & -1897.1 &  1.9\\
\enddata
\tablenotetext{a}{This observation was discarded, as discussed in Section \ref{sec:harps}}
\end{deluxetable*}

\end{document}